\documentclass[aps,prl]{revtex4}

\usepackage{graphicx}
\usepackage{times}

\begin{document}

\newcommand{\ee}{\end{eqnarray}}
\newcommand {\be}[1]{\begin{eqnarray} \mbox{$\label{#1}$}  }
\newcommand{\s}{{\cal S}}
\newcommand{\p}{\partial}
\newcommand{\pref}[1]{(\ref{#1})}

\newcommand\ie {{\it i.e.}, }
\newcommand\eg {{\it e.g. }}
\newcommand\etc{{\it etc. }}
\newcommand\cf{{\it c.f.}}
\newcommand\grad{\vec\nabla}
\newcommand\half{\frac 1 2 }
\newcommand\ls{\lambda^* }
\newcommand\e {\epsilon}
\newcommand{\nn}{\nonumber\\}

\newcommand{\ket}[1]{|#1\rangle}
\newcommand{\bra}[1]{\langle #1 |}
\newcommand{\braket}[1]{\langle #1 \rangle}

\newcommand{\dd}[2]{{\rmd{#1}\over\rmd{#2}}}
\newcommand{\pdd}[2]{{\partial{#1}\over\partial{#2}}}
\newcommand{\pa}[1]{\partial_{#1}}




\title{Effective photon-photon interaction in a \\
two-dimensional ``photon fluid''}

\affiliation {Department of Physics, University of California
Berkeley, California 94720-7300}
\affiliation {Physics Department, Stockholm University, SCFAB,
SE-10691 Stockholm, Sweden}

\affiliation {Department of Physics, University of Oslo, P.O. Box 1048
Blindern,
N-0316 Oslo, Norway}
\author{R.Y. Chiao$^{a}$, T.H. Hansson$^{b}$, J.M. Leinaas$^{a,c}$ and S.
Viefers$^{c}$}
\affiliation{${(a)}$Department of Physics,University of California
at Berkeley, Berkeley, CA 94720-7300, USA}\affiliation{${(b)}$
Stockholms universitet,
AlbaNova universitetscentrum,
Fysikum, SE - 106 91 Stockholm,
Sweden}\affiliation{${(c)}$Department of Physics,University of Oslo, P.O.
Box 1048 Blindern, 0316 Oslo, Norway}

\date{July 11, 2003}


\begin{abstract} We formulate an effective theory for the atom-mediated
photon-photon interactions in a two-dimensional ``photon fluid'' confined in
a Fabry-Perot resonator. With the atoms modelled by a collection of
anharmonic Lorentz oscillators, the effective interaction is
evaluated to second
order in the coupling constant (the anharmonicity parameter). The interaction
has the form of a renormalized two-dimensional delta-function potential,
with the renormalization scale determined by the physical parameters of the
system, such as density of atoms and the detuning of the photons
relative to the resonance frequency of the atoms.
For realistic values of the parameters, the perturbation series has
to be resummed, and the effective interaction
becomes independent of the ``bare'' strength of the anharmonic
term. The resulting expression for the non-linear Kerr susceptibility,
is parametrically equal to the one found earlier for a dilute gas of
two-level atoms.
Using our result for the effective interaction parameter,
we  derive  conditions for the formation of a
photon fluid, both for Rydberg atoms in a microwave cavity
and for alkali atoms in an optical cavity.
\end{abstract}

\maketitle

\section{Introduction}
\noindent

Quantum physics in two-dimensions has many interesting features
which give rise to effects that cannot be seen in
three-dimensional systems.
One of the most interesting
two-dimensional effects is the formation of incompressible electron fluids
that characterize the plateau states of the quantum Hall effect
\cite{Girvin90}.
Also high temperature superconductivity is believed to be essentially a
two-dimensional effect.

The interest in the physics of low-dimensional systems has motivated
both theoretical and experimental searches for new
kinds of two-dimensional many-body systems. In the case of weakly
interacting Bose-condensed atomic gases, two-dimensionality can be
reached in highly asymmetric traps \cite{asymmetric}, and quantum states
similar to the quantum Hall states have been predicted for such systems when in
rapid rotation \cite{BEC}.

Another idea that has been advocated by one of us \cite{chiao1,Chiao2000} is
that photons also, under specific conditions in photonic traps, can form a
two-dimensional system of weakly interacting particles with an
effective mass determined by the (fixed) momentum in the suppressed
dimension. Such a photon gas can in principle undergo phase transitions,
much like a cold atomic gas, and can in a condensed phase sustain
vortices and sound excitations, in a manner similar to that of an ordinary
superfluid.

This picture of the photons as a two-dimensional fluid has been based on the
(effective) Maxwell theory of electromagnetic waves in a
non-linear medium where only one longitudinal mode inside a cavity
is excited by an incoming laser beam. The corresponding mean field
equation has the same form as the Gross-Pitaevski equation, or the non-linear
Schr{\"o}dinger equation with a quartic non-linearity, and when coupled to an
external driving field it has been referred to as the
Luigiato-Lefever (LL) equation. The LL equation has been used to discuss the
apparence of transverse patterns in the light trapped inside
Fabry-Perot and ring
cavities
\cite{Lugiato87}.

The LL equation is  a non-linear classical
field equation,
but it can also be interpreted as a quantum field theory with the
electromagnetic field as an operator field. The non-linear term
is then viewed as a short range ($\delta$-function)
  photon-photon interaction.
This interpretation is the basis
for the photon fluid idea, and it has implications beyond the
classical non-linear optics description.

However, the interpretation of the non-linear field equation as a
quantum theory raises several questions. One   has to do with
the dimensional reduction itself. When only the fundamental
longitudinal mode is excited there is clearly an effective reduction of
dimension, since the dynamics is restricted to the two transverse
directions. This
corresponds to the situation where the cavity is small, with a length in the
longitudinal direction of the order of half a wave length. In the
optical regime
such a resonator is extremely small, and even if it can be made 
in principle, a
simpler realization of a small resonator is in the microwave
regime in conjunction with Rydberg atoms which can couple strongly to the
microwave photons. For the cavities that are presently used in laser
experiments the longitudinal mode is highly excited relative
to the fundamental mode, and in
this case two-dimensionality is obtained only as long as the
scattering to other longitudinal modes can be neglected.

Another question concerns the photon-photon scattering \cite{chiao2}. A
two-dimensional delta-function interaction is only well defined to lowest
order in perturbation theory, and in a full quantum description such a
short range interaction is meaningful only as a renormalized interaction.
This implies that the scattering amplitude is determined by a
renormalization length in addition to the interaction strength. In the
effective photon theory this is not a free parameter, but should be
determined by the full microscopic theory of the photons interacting
with the atoms
of the non-linear medium.

In this work we will address the first question simply by assuming that
only one longitudinal mode is excited.
Our main objective will then be to examine the photon-photon interaction
from a microscopic point of view. Of particular interest is to examine in
what sense the effective interaction  can be interpreted as a
delta function interaction and to determine how the renormalized strength of
the interaction depends on the physical parameters of the system.

The approach we will take is to derive the effective photon theory from
from the full quantum theory of the electromagnetic field and the
non-linear medium, rather than from a macroscopic description of the
electromagnetic field. However, we will use the simplified model of the
atoms in the medium as a collection of Lorentz oscillators supplemented
by a quartic oscillator term to account for the non-linearity \cite{Boyd}. At
the quantum level, the linear Lorentz oscillator model yields
``polaritons" as the
coupled atom-photon degrees of freedom, as shown by Hopfield and others
\cite{Hopfield58,Huttner92}.

In the next section (2) we use the Feynman path-integral method to find an
expression for the interaction between the physical modes of the coupled
photon-oscillator system in terms of an effective photon action.
In Section 3 we consider the effective theory for the low-lying transverse
momentum modes in a cavity where only the lowest longitudinal mode
is excited, and derive the corresponding two-dimensional low-energy effective
action. In the following section (4), we summarize the question of how to
correctly describe the renormalized delta function interaction in two
dimensions.
Then in Section 5, we relate this to an evaluation of quantum
corrections to
the effective interaction (to second order in the coupling parameter)
and determine
the leading logarithmic corrections to the scattering amplitude. In
Section 6  we
summarize the physical scales and discuss the conditions under which
interesting
quantum  phenomena like Bose-Einstein condensation and the formation
of two-photon
bound states may take place. In section 7 we examine two possible
scenarios for experimental realizations of a 2D photon fluid. Based on
order of magnitude estimates, we discuss under what conditions a photon
fluid in thermal equilibrium may form for millimeter-wave photons
interacting with Rydberg atoms and for optical photons interacting with
alkali atoms. Both cases might offer possibilities to observe
genuine quantum effects in an interacting photon gas, although our
estimates indicate severe constraints on the
physical parameters. We close this section by discussing a third possible
scenario, in which a
2D photon fluid forms just inside the surface of a high-Q microsphere of glass.
Concluding remarks are found in Section 8.

\section{2 The effective photon action}
\label{s2}
\noindent

For photons interacting off resonance with the atoms (\ie the oscillators),
the atoms have two types of effects on the scattered photons. There is a
linear effect, where the elastic scattering off the atoms changes the
dispersion of the photons. For photons with low transverse momentum in the
cavity, this leads to a renormalization of the effective mass of the
photons, which without the atoms is determined by the (fixed) longitudinal
photon momentum $\hbar k_L$, as
$m_0=\hbar k_L/c$. The other effect
is the non-linear or anharmonic effect of the photon-atom interaction,
which gives rise to the effective photon-photon interaction.

There are several ways to derive the effective
photon action from the full quantum theory of photons and oscillators,
all based on the assumption that the non-linear term is small and
can be treated as a perturbation. One way is to solve, as the first step,
the linear part of the problem exactly by diagonalizing the Hamiltonian.
As the next step the non-linear terms can be expressed in terms of the
transformed, decoupled variables; in this way the resonance
problems, which would appear in a more direct perturbative treatment, are
avoided. However, another simpler approach which we will adopt here, is
to derive the effective action of the electromagnetic field by use of the
Feynman path integral method, where a (non-local) field
transformation yields the result
without any matrix diagonalization. To check the result
of this method we have also performed, in Appendix A, 
a decoupling of the linear variables
by matrix diagonalization, and show that this can be done
in a way that is substantially simpler than the standard method.

In this section we do not impose the cavity boundary conditions, which
later will be used as constraints on the effective photon modes in order to
derive a dimensionally reduced theory.

We start from the total classical action
\be{act}
\s [\vec A, \vec R_{i}] = \s_{\gamma}[\vec A] + \s_{M}[\vec R_{i}] +
\s_{I}[\vec A, \vec R_{i}]
\ee
which is the sum of a free photon part $\s_{\gamma}$, a matter part
$\s_M$, for the atoms, and an interaction $\s_{I}$ which describes the
their coupling to the radiation. $\vec R_{i}$ describes the atomic
degrees of freedom, which here are given as the displacement vectors of a
discrete set of oscillators labeled by
$i$. In the following we will put $\hbar = c =1$.

The free photon part, $\s_{\gamma}$, is given by the Maxwell term
\be{S0}
{\cal S}_\gamma = \half \int d^3r\, dt\,   [E^2 - B^2] \text{ ,}
\ee
where $\vec E = -\dot{\vec A}$ and $\vec B = \nabla \times \vec A$ with
$\vec A$ satisfying
the Coulomb gauge condition $\nabla \cdot \vec A =0$. The oscillator
part includes an anharmonic term,
\be{SM}
{\cal S}_M =
   \int dt\, \sum_{i} \left(\frac{M}{2}{\dot{\vec R_i}}^2
       -\frac{M}{2}\omega_0^2 {\vec R_i}^2
       -\frac{\Lambda M^2}{4}{\vec R_i}^4
\right) \, ,
\ee
and the atoms are coupled to the
electromagnetic field via a dipole interaction,
\be{dip}
\s_{I} = \int\, dt\sum_i q\vec A(\vec r_{i})\cdot \dot{\vec R_i} \, .
\ee
Here $\vec r_{i} $ is the spatial
position of the i:th oscillator,
and we shall furthermore assume that the atoms are
uniformly distributed
in space with a number density $\rho$, so that discrete sums can be replaced
by integrals over a continuous position vector $\vec r$.
Since only the transverse part of $\vec R$ couples to the photon field,
it is consistent to neglect the longitudinal part and impose a
Coulomb ``gauge" condition
also on the oscillator, \ie  $\nabla \cdot \vec R =0$.

The effective photon action $\s_{eff}[\vec A]$ is defined by the
following (path) integral over the oscillator variables $R_{i}$
\be{effa}
e^{i\s_{eff}[\vec A]} = \int\prod_{i}{\cal D} R_{i}\, e^{i\s [\vec
A, \vec R_{i}]} \, .
\ee
To perform the $R_{i}$ integrations, we go to Fourier space and expand
to lowest order in $\Lambda$,
\be{genf}
e^{i\s_{eff}}
   &=& e^{i\s_\gamma}\int\prod_{i}{\cal D} R_{i}\, e^{i\int
\frac {d\omega} {2\pi} \sum_{i}  [ \frac M 2 \vec
R_{i}(-\omega)(\omega^{2}-\omega_{0}^{2})
\vec R_{i}(\omega) - i \omega q\vec A_{i}(-\omega)\cdot\vec R_{i}(\omega)]} \\
&[&1 - \frac {i\Lambda M^2} {4} \int \prod_{n=1}^{4} \frac {d\omega_{n}}
{2\pi}
\, 2\pi\delta(\sum_{n=1}^{4}
\omega_{n}) \vec R_{i}(\omega_{1})\cdot  \vec R_{i}(\omega_{2})
\vec R_{i}(\omega_{3})\cdot  \vec R_{i}(\omega_{4}) + O(\Lambda^{2}) ]
\nonumber
\, .
\ee
The first term is evaluated directly by completing the square and
making the shift \\
$\vec R_{i}(\omega) \rightarrow \vec R_{i}(\omega)
- i \omega \frac q M \vec A_{i}{\cal D}_{0}(\omega)$, with ${\cal
D}_{0}(\omega)$  as the retarded propagator,
\be{prop}
{\cal D}_{0}(\omega) = \frac 1 { \omega^{2} - \omega_{0}^{2} +
i\omega \epsilon } \, .
\ee
  It yields the quadratic
part of the effective action,
\be{eff2}
{\cal S}_{eff}^{(2)} &=& \frac 1 2  \int d^{3}x\int \frac {d\omega} {2\pi}
\, \vec A (\vec x, -\omega) \left[ \omega^{2} + \nabla^{2}
        - \eta^2\omega^2  {\cal D}_{0}(\omega) \right] \vec A (\vec x, \omega)
\text{,}
\ee
where in this expression we have taken the continuum limit $\sum_i
\rightarrow \rho\int d^3x$ and also introduced the effective plasma frequency
$\eta= (\rho q^2/M)^{1/2}$.

The quartic $\vec R$ term in \pref{genf} is most easily evaluated by
again performing the shift in $\vec R_{i}(\omega)$, to give $\vec
A$ dependent terms of the form $ \vec A^{4}$, $ (\vec
A\cdot\vec A)( \vec R\cdot
\vec R$) and $(\vec A\cdot\vec R )(\vec A\cdot\vec R) $. The  $ \vec
A^{4}$ term can be directly  re-exponentiated and gives a quartic
contribution to the effective action, which in the continuum limit is
\be{eff4}
{\cal S}_{eff}^{(4)} = -   \frac
{2\pi \eta^4\Lambda} {4 \rho}
       \int d^{3}x \int  \prod_{n=1}^{4} \left[
      \frac {d\omega_{n}}{2\pi}\omega_{n} \, { \cal D}_{0}(\omega_{n})\right]
      \delta(\sum_{n=1}^{4}\omega_{n})
  \vec A(\omega_{1},\vec x)\cdot \vec A(\omega_{2},\vec x)\,
     \vec A(\omega_{3},\vec x)\cdot \vec A(\omega_{4},\vec x) \, .
\ee
The terms proportional to $ (\vec A\cdot\vec A)( \vec R\cdot
\vec R$) and $(\vec A\cdot\vec R )(\vec A\cdot\vec R) $ can be evaluated
using the formula,
\be{form}
\int\prod_{i}{\cal D} R_{i}\, e^{i\int \frac {d\omega} {2\pi} \sum_{i}
[ \frac M 2 \vec R_{i}(-\omega)(\omega^{2}-\omega_{0}^{2})
\vec R_{i}(\omega) } R_{j}^{a}(\omega_1)R_{k}^{b}(\omega_2)
= {\cal N} \frac {2\pi i} M {\cal D}_{0}(\omega_1)
\delta (\omega_{1} + \omega_{2} ) \delta_{ab}  \delta_{jk}\,\,\,,
\ee
where ${\cal N}$ is a field-independent normalization factor. They give,
in principle, a correction term to the quadratic action, 
but due to the integration
over ${\cal D}_{0}(\omega_1)$ this correction term vanishes.

Note that the expansion and re-exponentiation of the non-linear term will
generate correction terms, but these are higher order in $\Lambda$ and
will be neglected. Thus, to first order in $\Lambda$, the effective action
is given by the quadratic part \pref{eff2} and the quartic term
\pref{eff4}.

The effective action defined by \pref{eff2} and \pref{eff4} corresponds
to a Lagrangian that is non-local in time. However by a further
transformation it can be brought into a local form. We first note that
the quadratic part of the action defines a modified dispersion equation
\be{disp}
\omega^{2} -k^2
     - \eta^2\frac {\omega^2}{\omega^2-\omega_0^2}=0
\ee
with solutions
\be{disp2}
\omega_\pm^{2}(k^2)=\half \left( k^2+\omega_0^2+\eta^2 \right) \pm
\half \sqrt {\left( k^2+\omega_0^2+\eta^2\right)^2-4\omega_0^2  k^2} \, .
\ee
This equation defines the dispersion of the ``polaritons", \ie the two
decoupled degrees of freedom of the linear problem which mixes the
photon and dipole variables. For $k^2<\omega_0^2+\eta^2$ $
\omega_-$ represents essentially the photon mode and $\omega_+$ the dipole
mode, whereas for $k^2>\omega_0^2+\eta^2$ the interpretation of the two
modes is reversed. In the intermediate interval with
$k^2\approx\omega_0^2$ the photon and the dipole modes are strongly mixed.

The following field transformation is now applied
\be{transf}
\vec A(\vec k, \omega)\rightarrow \vec A^{(\pm)}(\vec k,
\omega)=\sqrt{\frac{\omega^2-\vec
k^2-\eta^2\frac{\omega^2}{\omega^2-\omega_0^2}}{\omega^2-
\omega_{\pm}(k^2)}}\;\; \vec A(\vec k, \omega)
\ee
and this gives for the quadratic part of the action
\be{eff2b}
S^{(2)}_{eff}=\half \int \frac{d^3k}{(2\pi)^3}\int
\frac{d\omega}{2\pi}\;\vec A^{(\pm)}(\vec k,
\omega)^*\;(\omega^2-\omega_{\pm}^2)\;\vec A^{(\pm)}(\vec k, \omega)\, .
\ee
The dependence on $\omega^2$ shows that the transformed action corresponds
to a Lagrangian that is local in time. The non-locality in time has been traded
for a non-locality in space, but this is less problematic in a Lagrangian
formulation. Note, however, the ambiguity in the transformations
\pref{transf}, depending of which one of the solutions
$\omega_{\pm}$ we choose. Clearly, the relevant choice is the one
which fits the energy of the photons in the effective theory. This means
$\omega_-$ for the case of red detuning (energy below $\omega_0$) and
$\omega_+$ for blue detuning (energy above $\omega_0$).

When the transformed field is introduced in the quartic part of the
effective action we make a further simplification by assuming that the
fields satisfy the dispersion equation of the linear problem. This
allows the following substitution
\be{subs}
i\omega{\cal D}_0(\omega)\vec A(\vec k,
\omega)&&\rightarrow \lim_{\omega^2\rightarrow\omega_{\pm}^2}\
\frac{1}{\omega^2-\omega_0^2}\sqrt\frac{\omega^2-\omega_{\pm}^2}
{\omega^2-k^2-\eta^2\frac{\omega_0^2}{\omega^2-\omega_0^2}} \;{ \vec
E}^{(\pm)} (\vec k, \omega) \nn
&&= \frac{1}{\sqrt{(\omega_{\pm}^2-\omega_0^2)^2+\eta^2\omega_0^2}}
\;{
\vec E}^{(\pm)}(\vec k, \omega) \, ,
\ee
where ${ \vec E}^{(\pm)} (\vec k, \omega)
= i\omega_{\pm} \vec A^{(\pm)}(\vec k, \omega) $.
For the interaction part of the action this gives the following
expression,
\be{eff4b}
{\cal S}_{eff}^{(4)} &&= -   \frac
{(2\pi)^4 \eta^4\Lambda} {4 \rho}
         \prod_{n=1}^{4}\left[ \int\frac
{d^{3}k_n\;d\omega_{n}}{(2\pi)^4}
\frac{1}{\sqrt{(\omega_{\pm}(k_n)^2-\omega_0^2)^2+\eta^2\omega_0^2}}
\right]
     \delta(\sum_{n=1}^{4}\vec k_{n}) \delta(\sum_{n=1}^{4}\omega_{n})
\nn
  &&\times \vec E^{(\pm)}(\vec k_1,\omega_{1})\cdot \vec
E^{(\pm)}(\vec k_2,\omega_{2})\,
     \vec E^{(\pm)}(\vec k_3,\omega_{3})\cdot \vec
E^{(\pm)}(\vec k_4,\omega_{4}) \, .
\ee
The application of the dispersion equation to the field variables of the
interaction term can be justified when this term is used perturbatively,
with the fields satisfying the field equation of the unperturbed system.
However, it is interesting to note that the expression
\pref{eff4b} in fact is valid beyond this approximation, as is
demonstrated by the diagonalization of the quadratic problem performed
in Appendix A.

\section {3  Dimensional reduction and the effective 2d $\phi^{4}$
theory}
\noindent
Due to the boundary conditions imposed by the mirrors, the component of
the photon momentum normal to the mirrors (the longitudinal momentum) is
quantized at discrete values. We assume an idealized situation with
infinite flat mirrors, thus the longitudinal momenta are quantized as
$k_{L}=\pi n/L$, with $L$ as the  distance between the mirrors and $n$ as
an integer.
We also assume photons to be fed to the cavity by a laser (or maser) tuned
close to resonance with one of the modes,  either slightly below (red
detuning), or slightly above the resonance (blue detuning). However,
we do not take
the effect of photons entering or departing the cavity explicitly into account,
and in this sense we consider an idealized situation with perfectly
reflecting mirrors. All the photons inside the cavity are assumed to be
trapped in the same longitudinal mode, and throughout the paper we will
assume this to be the lowest  mode
$n=1$.

The transverse components of the photon momentum $\vec k$ we assume to be
restricted to small values, $k_T<<k_L$. The dispersion of free
(non-interacting)
photons inside the Fabry-Perot resonator becomes essentially that of 2D
massive, non-relativistic particles \cite{chiao1},
\be{dispx}
\omega_k = \sqrt{k_T^2+k_L^2} \approx k_L + \frac{k_T^2}{2k_L},
\ee
with the longitudinal momentum $k_L$ playing the role of the photon mass.
The dimensional reduction is then based on the assumption that only one
longitudinal mode (the lowest) is excited, and that scattering to other
modes can be neglected.
We should stress that this does not mean that higher modes are not
important as virtual states in the perturbative expansion - in fact they are.

For simplicity we shall in the following refer to the transverse momentum
simply as $k_T\equiv k$ and the longitudinal momentum as $k_L=\pi/L\equiv m_0$.

When the coupling between the photons and the oscillators
is taken into account, the dispersion equation is
given by \pref{disp2}, and the relation between $\vec k$ and $\omega(k)$ is no
longer so simple. In Fig.1a  $\omega_-(k)$ and $\omega_+(k)$ are shown as 
functions of the transverse momentum for red detuning, which is the case we
will first consider. It displays how for small momenta,
$\omega_-(k)$ corresponds  to the ``photon branch" with quadratic dependence
on $k$, while for large momenta the photon branch is represented by
$\omega_+(k)$ .
In the following we will simply refer to excitations of this mode as
``photons"  and the other one as ``dipoles". Due to the mixing there is an
avoided level crossing at intermediate momenta, where there is no clear
distinction between the photon and the dipole mode.
\begin{figure}
\includegraphics[width=15cm,height=6cm]{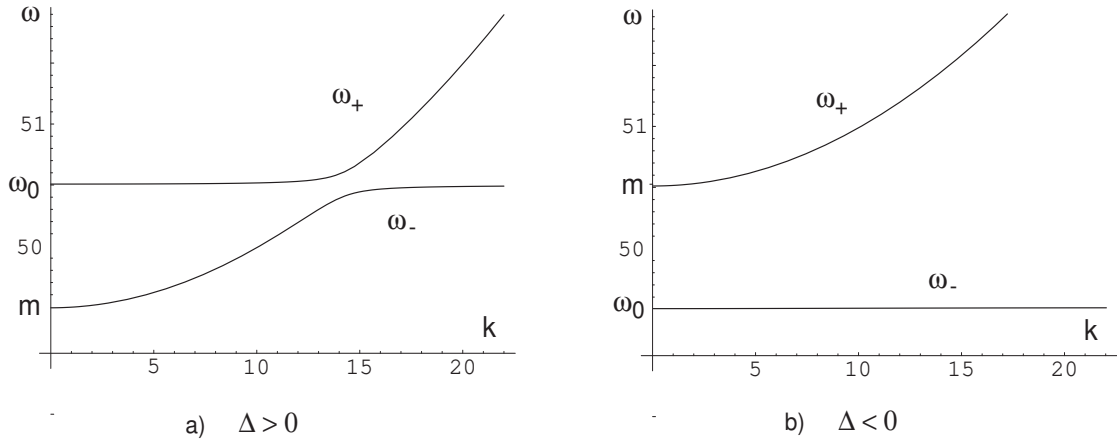}
\caption{Dispersion curves for the photon like and dipole like excitations.
Fig 1a shows the curves for red detuning ($\Delta=\omega_0-m>0$), where the
photon branch corresponds to the curve $\omega_-$ for low momentum $k$ and to
$\omega_+$ for high momentum. Fig 1b shows the curves for blue detuning
($\Delta=\omega_0-m<0$), where the photon branch corresponds to $\omega_+$ for
all momenta. The curves are shown in dimensionless units with $|\Delta|=1$. The
parameter values are
$\omega_0=100$ and $\eta=0.1$.}
\end{figure}

For the case of interacting photons, just as for non-interacting photons,
a low-momentum description can be made where the photons appear as
non-relativistic, massive particles. Thus, when the photon frequency is
separated from the resonance frequency of the oscillator mode by a detuning
gap
$\Delta = \omega_0-m_0$, and the transverse momentum is restricted by
\be{scales}
k^2 \ll m_0 \Delta \ll \omega_0^2 \  \, ,
\ee
then the previous expressions for $S_{eff}^{(2)}$ and $S_{eff}^{(4)}$
(\pref{eff2} and \pref{eff4}) define a low momentum effective action for the
photons, with a  dispersion relation of a non-relativistic form
\be{sol}
\omega(k) = m + \frac{k^{2}}{2m} + ...  \, .
\ee
In the following we shall in addition assume weak coupling
between the photons and dipoles, in the sense
\be{weakcoup}
\eta^2  \ll \Delta^{2}  \ .
\ee
The effective photon mass is then given by
\be{effmass}
m^2=m_0^2- \eta^2 \frac{m_0}{2\Delta} + ... \, ,
\ee
with a small renormalization of the mass due to the interaction with the dipole
field. Note that the weak coupling condition \pref{weakcoup} is not
essential for
the non-relativistic description of the 2D photons, but is introduced
to simplify
the calculations. In physical realizations of the 2D photon gas one
may also have
to consider the case of strong mixing of the photon and dipole
degrees of freedom,
as discussed in the section on Rydberg atoms below.

With the longitudinal momentum fixed to $m_0$ and $\omega(k)$ approximated by
$m$, the field variable
$A^{(-)}$ can be written as,
\be{A}
\vec A^{(-)}(\vec r,z,t)=  \sum_{a=\pm 1} \int\frac{d^2k}{(2\pi)^2}
              \frac 1 {\sqrt{m L}}
          \sin(m_0 z)\left[   e^{-i(mt-\vec k\cdot\vec
r)} \;\vec\epsilon_{a}(\vec k) \,
                        \phi_{a}(\vec k, t) + c.c. \right] \, ,
\ee
where $ \vec\epsilon_{a}(\vec k)$ are the polarization vectors and
$\phi_{a}(\vec k, t)$ are the corresponding field components, which now only
depend on the transverse momentum $\vec k$.  Note that both the
frequency and the momentum of the longitudinal mode have been extracted
from
$\phi_{a}(\vec k, t)$ in order to express this as a slowly varying field.

When the assumptions about small transverse momentum and weak coupling are
imposed, the quadratic part of the effective action  gets the
form (to order $k^2$),
\be{phieff2}
{\cal S}_{\phi}^{(2)} = \int d^2r dt  \,
\sum_{a=\pm 1}
      \left[ \frac{i}{2}\left(\phi_a^* \dot\phi_a
          - \dot\phi_a^*\phi_a \right)
      -\frac{1}{2m} \left| \vec\nabla\phi_a \right|^2 \right]
\ee
Here we have neglected terms proportional to $ \dot\phi/m$
(slowly-varying field
approximation), and $\vec\nabla$ is now the two-dimensional gradient.
Note that the
two polarization directions appear as two species of particles. The action has
the standard form of a non-relativistic, free field theory.

We will now consider the interaction term. In the same approximation as
used above we have
\be{approx}
\frac{1}{\sqrt{(\omega_{-}(k)^2-\omega_0^2)^2+\eta^2\omega_0^2}}
\approx \frac {1} {2m|\Delta|}\, .
\ee
If this $k$-independent expression is used in the interaction part of
the action,
\pref{eff4}, and the $\vec E^{(-)}$ field is
expressed in terms of $\phi_a$, it simplifies to
\be{phieff4}
{\cal S}_{\phi}^{(4)} = - \frac {\pi g} {2m}
\int d^2r dt  \, \sum_{a=\pm 1} \left[|\phi_{a}|^{4} +
           2|\phi_{a}|^{2}|\phi_{-a}|^{2}\right] \, .
\ee
where
$g$ is the (bare) interaction strength given by
\be{inst}
g = \frac 1 {(2\pi)^2}\frac{3}{16} \frac \Lambda {\rho}
\left(\frac {\eta} {\Delta}\right)^4 \, .
\ee

The first term of \pref{phieff4} can
be interpreted as a delta function interaction between photons with the same
helicity, the second between photons of opposite helicities. In the simplest
case,  with only one type of photon polarization ($\phi_a =\phi$), the
corresponding interaction Lagrangian simplifies to
\be{iden}
{\cal L}_{\phi}^{(4)} = - \frac {\pi g} {2m}
|\phi(\vec r,t)|^{4} \, .
\ee
Thus, with the approximations used, we reach a form of the effective photon
Lagrangian which agrees with the nonlinear Schr\"odinger equation previously
derived from the classical field theory of a dimensionally reduced
Maxwell field interacting with a non-linear medium \cite{Akhmanov}. The photons
behave like 2D massive particles with a repulsive, pointlike interaction.

In the case of blue detuning, \ie $\Delta < 0$, the situation is quite
different. Typical dispersion curves are shown in Fig.1b, and we see that
now the $\omega_+(k)$ branch is photon-like both for small and large momenta,
and there is no avoided level crossing, but only a level repulsion at small
momenta.

However, also for blue detuning a low momentum effective action can be given
where all the above formula, derived for red detuning, still hold for
scattering
between the  photons. But an important change is that the photons now are
the ``high energy" particles relative to the dipoles, and it is energetically
possible for them to ``decay" into the low lying dipole modes. At  tree level
(to first order in the interaction) the dangerous process is
$\gamma
\gamma
\rightarrow \gamma {\rm D}$ which for blue detuning can conserve both energy
and momentum.  The corresponding interaction piece in the Lagrangian can again
be read from
\pref{eff4b},  but now with one of the fields as $\vec E^{(-)}$ (dipole field)
and the three others as $\vec E^{(+)}$ (photon field). In the low momentum
approximation it is
\be{mixing}
{\cal L}_{\rm mix}^{(4)} = - \frac {\pi g} {m} \frac {\Delta} {\eta}
(\tilde\phi^*\phi)(\phi^*\phi)  \, ,
\ee
where $\phi$ is the photon field and $\tilde\phi$ is the dipole field.

An important quantity for assessing whether a gas of blue
detuned photons
can be maintained in the cavity, is the ratio between the cross section of
the above (`inelastic') decay process, 
and the normal (`elastic') scattering 
$\gamma \gamma \rightarrow \gamma \gamma$ 
induced by the interaction \pref{inst}. A straightforward
calculation yields,
\be{cross}
\sigma_{tot}^{el.}(k) &=& \frac{1}{2}\left(\pi g\right)^2 \frac 1 k   \\
\sigma_{tot}^{inel.}(k) &=&  \left(\frac {\pi g\Delta} \eta \right)^{2}
\sqrt{\frac 2 {m| \Delta |} }   \nonumber
\ee
where $k$ is the momentum in the center of mass. 
Thus, demanding $\sigma_{tot}^{inel.} \ll \sigma_{tot}^{el.} $, 
implies the condition 
\be{decaycond}
k \ll \left( \frac{\eta}{\Delta} \right)^2 \sqrt{ \frac{m|\Delta |}{8} }
\, .
\ee
We shall return to numerical estimates of the physical parameters
in section 7.

We conclude this section with some comments on Galilean invariance.
The effective action, given by \pref{eff2} and \pref{eff4}, is neither Lorentz
nor Galilean  invariant, since the relativistic photons are coupled to dipoles
defined in a fixed frame. Nevertheless the effective theory defined by
\pref{phieff2} and \pref{phieff4} is Galilean invariant due to the low
momentum  approximations made in the vertices. In the next section we will
consider loop effects where the low momentum approximations are not any longer
valid, and it is far from obvious that the resulting corrections to the
effective theory will respect the Galilean invariance. As we shall see,
however, the leading corrections do have this symmetry, so the interpretation
of the photons as a nonrelativistic Bose system has validity beyond
the Born approximation.

\section{4  Renormalization of the $\delta$-function interaction. }

In a many-particle interpretation the interaction Lagrangian \pref{iden}
corresponds to a delta-function potential
\be{deltaf}
V(r)=       \frac {\pi g} m  \sum_{i<j}\delta^{(2)}(\vec r_{ij})
\ee
where $\vec r_{ij}$ is the two-particle relative position.
However, it is well known  that a pure delta function
interaction in dimensions higher than one is not well defined beyond first
order in perturbation  theory. In two dimensions the second order term
gives rise to a logarithmic divergence in the  scattering amplitude. To make
the delta function interaction meaningful requires regularization of the
interaction and renormalization of the interaction strength. As discussed in
ref.\cite{jackiw1} the form of the s-wave scattering amplitude for such a
renormalized interaction is
\be{sca}
f(k) = -(2\pi)^2\sqrt{ \frac 1{2\pi k}  }\,\frac {\frac {g_r} 4} {1 -
\frac {g_r}{4}
\left( \ln \frac {k^{2}} {\mu_r^{2}} -i\pi\right) }
\ee
where $g_r$ is the renormalized coupling constant and $\mu_r$ is a
new parameter
that is introduced by the renormalization (the renormalization scale). The
corresponding phase shift for small $g_r$ is given by $\delta_0 = \pi g_r/4$,
which is the Born approximation value when $g=g_r$. For small valus of
$k$ it approaches the universal expression,  $\delta_0 \approx \pi/\ln k^2 $,
that  is common  for a large class of short range potentials \cite{chadan98}.

Formally, the delta function interaction in two dimensions is
dimensionless, \ie
it scales as the kinetic energy. However, the renormalization breaks the
scaling symmetry and introduces a length scale through the parameter
$\mu_r$. This
is similar to the situation in QCD, where the effect is referred to as
dimensional transmutation \cite{thorn}. One should note that the two
parameters $g_r$ and
$\mu_r$ are not independent. Thus, $g_r$ may be fixed as the bare parameter $g$
and all the effect of the renormalization may be absorbed in $\mu_r$, or $g_r$
may be viewed as depending on $\mu_r$, where $\mu_r$ is chosen to match the
physical momentum interval. In the latter case $g_r$ is referred to as an
effective (or running) coupling constant. The explicit dependence on $\mu_r$
is given by
\be{run}
{\frac 1 g_r} = {\frac 1 g} - \ln{\frac {\mu_r^2} {\mu_0^2}}
\ee
with $\mu_0$ as a constant.
 From this expression we notice that for large
momenta (large $\mu_r$) the effective coupling constant goes to zero,
so this is a
quantum mechanical analogue of the asymptotic freedom of QCD. Note
also the curious fact that for sufficiently large $k$ the effective interaction
is always attractive, irrespectively of the sign of the bare parameter $g$,
whereas it in the other limit is repulsive.

The above discussion refers to a situation where
the $\phi ^4$ theory is treated as a fundamental
theory where $g_r$ and
$\mu_r$ are free parameters, to be determined by experiment.
However, treated as a an effective (low energy) theory they are
in principle determined by the
physical parameters of the complete system. For example, in the case of a quasi
two-dimensional atomic Bose gas in a highly asymmetric trap, the
renormalization scale of the two-dimensional theory is essentially given by the
extension of the trap perpendicular to the plane in which the atoms
move \cite{petrov,hansson1}.

In the present case the renormalized interaction strength may be determined by
taking more explicitly into account the effect of the dipole degrees of
freedom. This we do by using Schr{\"o}dinger perturbation thery,
with the interaction
Hamiltonian extracted from the the action \pref{eff4b}, to
calculate contributions to the scattering amplitude
beyond the Born approximation. With the expression for the scattering
amplitude given by \pref{sca}, which we assume to be correct for low momenta
$k$, we note that the  renormalization scale $\mu_r$ can be determined from the
contribution to second order in $g$ (with $g_r=g$). In such a second
order calculation of the scattering amplitude the contributions from
the dipole mode cannot be neglected, since the intermediate
states are not restricted to low momenta. Thus,  both the field modes
$\vec E^{(\pm)}$ are included in the calculation, and  the exact expressions
for the mode frequences
$\omega_{\pm}(k)$ are used rather than the low momentum
approximations.

In the following we present a simple calculation of the leading
contributions to the scattering amplitude. The result shows that the form of
the amplitude is as expected and it gives an estimate of the renormalization
scale  $\mu_r$.
In Appendix B we perform a more complete calculation of some of
the  non-leading terms in the scattering amplitudes and give the corresponding
expressions for the scattering amplitude, for both red and blue detuning.

\section{5 The scattering amplitude to $O(g^2)$. }

Consider the T-matrix, related to the scattering matrix by
$S_{fi}=\delta_{fi}-2\pi i\delta(E_f-E_i)  T_{fi}$, where $E_f$ and
$E_i$ are the
energies of the final and initial states. To second order $T_{fi}$ has
the form
\be{T}
T_{fi}=\langle
k'_1k'_2|H_{int}|k_1k_2\rangle+\half \int d^2p_1d^2p_2 \frac{\langle
k'_1k'_2|H_{int}|p_1p_2\rangle\langle
p_1p_2|H_{int}|k_1k_2\rangle}{\omega(k_1)+\omega(k_2)-\omega(p_1)-\omega(p_2)+i\
e}
\ee
where we here have simplified the notation by the sum over field modes, 
longitudinal momenta and polarization
variables in the intermediate state.
The form of the interaction matrix element is
\be{intmat}
\langle
p_1p_2|&H_{int}&|k_1k_2\rangle=\frac{1}{(2\pi)^3}\frac{\Lambda \eta^4}{2\rho
\Delta^2}\delta(k_1+k_2-p_1-p_2)(\delta_{n_1n_2}+\half\delta_{n_11}\delta_{n_2
1}-\delta_{n_1n_2-2})\nn
&\times&
\sqrt{\frac{\omega(p_1)}{(\omega(p_1)^2-\omega_0^2)^2+\eta^2\omega_0^2}
\;\frac{\omega(p_2)}{(\omega(p_2)^2-\omega_0^2)^2+\eta^2\omega_0^2}}  \;\vec
e^{\;*}  \cdot
\vec e_{\alpha}(\vec p_1)\; \;\vec e^{\;*}  \cdot \vec
e_{\beta}(\vec p_2)
\ee
where we have approximated the energy of the incoming photons $(k_1$ and $k_2$)
by $m$, since these are in the lowest longitudinal mode. These
photons have the same
polarization vector $\vec e$, while the particles
(``polaritons") in the intermediate state have polarization vectors
$\vec e_{\alpha}(\vec p_1)$ and
$\vec e_{\beta}(\vec p_2)$. The quantum numbers $n_1$ and $n_2$ determine the
longitudinal momenta of the particles in the intermediate state.

In the low-momentum approximation the T-matrix element is independent
of the sum of
the transverse momenta, $\vec K=\vec k_1+\vec k_2$, \ie it is Galilean
invariant. This follows since $\omega(k_1)$ and $\omega(k_2)$
can be approximated by
$m$ in all places, except in the energy denominator when the
intermediate particles
are also in the lowest longitudinal mode. In that case the pole at
$p_1^2+p_2^2=k_1^2+k_2^2$ makes the $k$-dependence important.
However, due to momentum
conservation, we have in the low-energy approximation
\be{galilei}
\omega(k_1)+\omega(k_2)-\omega(p_1)-\omega(p_2)&=&\frac{1}{2m}(k_1^2+k_2^2-p_1^2
-p_2^2)\nn
&=& \frac{1}{2m}(2k^2-2(\vec p_1-\half\vec K)^2)
\ee
where $\vec k=\half(\vec k_1-\vec k_2)$ is the relative momentum. The
expression shows that
$\vec K$ can be absorbed in the integration variable $\vec p_1$.

Thus, the T-matrix has the following momentum dependence
\be{Tmat}
T_{fi}=\delta(\vec K_f-\vec K_i)\,{\cal T}(\vec k_f,\vec k_i)
\ee
where $\vec K_f$ and $\vec K_i$ are the sum of momenta for the outgoing and
incoming particles and
$\vec k_f$ and $\vec k_i$ are the relative momenta. For pure s-wave
scattering (delta function interaction) the reduced T-matrix element ${\cal
T}$ only depends on the magnitude of the relative momentum,
\be{calT}
{\cal T}(\vec k_f,\vec k_i)={\cal T}(k^2)
\ee
and is related to the s-wave scattering amplitude through
\be{samp}
f(k^2)=-(2\pi)^2m\sqrt{\frac 1 {2\pi k}} {\cal T}(k^2)
\ee

Since we are interested in the behaviour of the scattering amplitude for small
transverse momenta, in the first order expression we
simply put them equal to zero. With all photons in the same helicity state,
the first order contribution is
\be{M1}
{\cal T}_1(k^2) = \frac {g} {2\pi  m}
\ee
in accordance with the expression for the low-energy Lagrangian \pref{iden}. To
second order there are three diagrams shown in fig. 2.
Diagram 2a corresponds to two particles in the intermediate states,
while  diagrams 2b and 2c correspond to four and six particles in the
intermediate
states.  Thus, the contributions from diagrams 2b and 2c are
suppressed by the
energy denominator, since the energy difference between the two initial and the
four or six intermediate particles necessarily has to be large on
the scale set by
the transverse momentum. For this reason we shall only consider contributions
from diagram 2a. Note that there are two types of particles in the
intermediate state, characterized by energies $\omega_-$ and
$\omega_+$. As an important
point also note that the transverse  momenta of the intermediate
particle states
cannot be assumed  to be small, and also excitations
to higher longitudinal momenta have to be considered.
We now discuss how to calculate the leading contributions to diagram 2a.
\begin{figure}
\includegraphics[width=12cm,height=2cm]{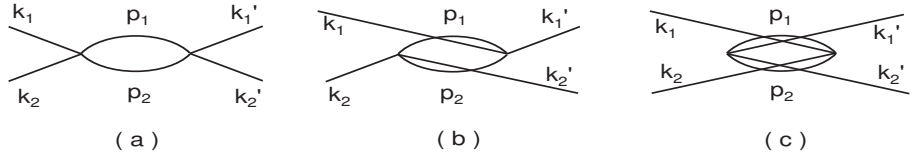}
\caption{Second order contributions to the scattering amplitude. Only
contributions from diagram (a) are included in this paper, since the
contributions
from (b) and (c) are suppressed by a factor $\Delta/m$.}
\end{figure}

At low momenta, there is a potentially large contribution to diagram
2a, when the energy denominator vanishes, and as expected this will give
rise to the logarithmically infrared divergent term $\sim \ln k^2$ in
\pref{sca}. This term is dominant in the limit of
asymptotically small momenta where the approximations leading to
\pref{eff4b} become exact.

The importance of the high momentum contribution comes from the fact that
$\omega_-$ does not increase with momentum, but rather approaches the resonance
value $\omega_0$, {\cf} Fig. 1. Thus, even if intermediate states with
frequencies close to
$\omega_0$ are considered as highly excited relative to the low energy photons,
the large number of dipoles may make contributions from these modes
important. In fact, if the dipoles are treated as a continuum, the integral
over intermediate momenta will diverge. In reality we know that there is a
physical cutoff related to the discreteness of the system of dipoles. We
introduce this simply as a cutoff in momentum at a value corresponding to the
(average)  distance between the dipoles.

With a clear separation of the scales in the momentum integrals the leading
contributions from high and low momenta can be estimated separately.
In order to see
how this works we examine the following toy problem.  Consider the integral
\be{toy}
I = \int_0^{\Lambda^2} dx\, \frac 1 {x+p^2} \frac {m^4 + x^2} {x + \Delta^2}
\ee
which can be evaluated exactly, to give
\be{toyex}
I =\Lambda^2+\frac{1}{{p^2}-{{\Delta }^2}}\Big({m^4}\log
\big[\frac{{p^2}({{\Delta }^2}+{{\Lambda }^2})}{{{\Delta }^2}({p^2}+{{\Lambda
}^2})}\big]+{p^4}\log \big[\frac{{p^2}}{{p^2}+{{\Lambda
}^2}}\big]-{\Delta^4}\log
\big[\frac{{\Delta^2}}{{\Delta^2}+{{\Lambda }^2}}\big]\Big)
\ee
However, assuming
\be{par}
p^2\ll \Delta^2 \ll m^2\ll \Lambda^2 \,
\ee
the integral can be estimated by the following approximation,
\be{toyap}
I &\approx& \int_0^{\Delta^2} dx\, \frac 1 {x+p^2} \frac {m^4 + x^2}
{x + \Delta^2}
+ \int_{m^2} ^{\Lambda^2} dx\, \frac 1 {x+p^2} \frac {m^4 + x^2} {x +
\Delta^2} \\
&\approx& \frac {m^2} {\Delta^2} \int_0^{\Delta^2} dx\, \frac 1 {x+p^2} +
\int_{m^2}^{\Lambda^2} dx\,1 \nn
&\approx& - \frac {m^4} {\Delta^2} \ln \frac {p^2}
{\Delta^2}  + \Lambda^2 \nonumber \, ,
\ee
where the low and high momentum contributions are treated separately.
This expression reproduces the exact result up to $O(m^2/\Lambda^2)$.
Below we will examine the leading contributions to the second order scattering
amplitude in this way, by evaluating separately the contributions
from low and high
momenta. In this calculation  the detuning parameter and the
photon mass will play the role of $\Delta$ and $m$ in the toy problem.  We
refer to Appendix B for a more complete treatment.

\subsection {The high-momentum contribution}

For large momenta the important contribution comes from the term with
two dipole excitations in the intermediate state. For these excitations we have
have $\omega(p)^2\approx\omega_0^2$ and the momentum integral is divergent
without the cutoff. The only effect of the coupling between the photons and the
oscillators appears in the denominators of the form
\be{osc}
\frac{1}{(\omega(p)^2-\omega_0^2)^2 + \eta^2 \omega_0^2}
\rightarrow \frac{1}{\eta^2 \omega_0^2}
\ee
where it prevents the expression from diverging when
$\omega(p)\rightarrow\omega_0^2$.

We can neglect
contributions  from the transverse momenta of the scattered (external) photons,
since these are smaller by
${\cal O}[ (k/m)^2]$ relative to the leading term. This means that
the contribution
gives rise to a
$k$-independent renormalization of the interaction strength. We also
neglect terms
that are higher order in the coupling strength $\eta$.
With $\omega(k)=m$ for the external photon states and
$\omega(p)=\omega_0$ for the intermediate states the energy denominator is
approximated by
\be{oscdenom}
\frac{1}{\omega(k_1)+\omega(k_2)-\omega(p_1)-\omega(p_2)}\rightarrow
-\frac{1}{4m\Delta}
\ee
and the matrix elements of the interaction get a simple form
\be{Hint}
\left\langle {p_1,p_2}
\right|H_{int }\left| {k_1\,k_2} \right\rangle &=&
\frac{\Lambda \eta^2 }{2(2\pi)^3 \rho \Delta^2 \omega_0} \;
\;\vec e^{\;*}  \cdot
\vec e_{\alpha}(\vec p_1)\; \;\vec e^{\;*}  \cdot \vec
e_{\beta}(\vec p_2)\nn
&\times& \delta(\vec p_1+\vec p_2 -
\vec k_1-\vec k_2)(\delta_{n_1 n_2}+\half \delta_{n_1 1}\delta_{n_2
1}-\delta_{n_1
(n_2-2)})
\ee
For high momenta the summation over the
polarization vectors gives  simply a factor
$3/4$  for each intermediate particle (as discussed in Appendix B)
and the momentum
integral and sum therefore gets trivial, with a momentum-independent matrix
element. Integrating over transverse momenta and summing over
longitudinal momenta
gives
\be{M2a}
{\cal T}_{2a}(k^2) =
-\frac{3\pi}{64}\frac{\Lambda}{m^2\Delta}{\cal T}_1
   \equiv  \alpha {\cal T}_1 \, ,
\ee
where
$\alpha$ is a new dimensionless parameter. Since the
dimensional reduction is not effective at high momenta (we have to
sum also over
the longitudinal momenta) this parameter is a characteristic of the full
three-dimensional theory, and is a measure of the importance of renormalization
of the nonlinear effects for intermediate momenta, \ie for $g \ln
k^2\Delta^2\ll
\alpha$.

In order to obtain the finite result \pref{M2a}
we  have introduced a cutoff, $p_{max} =\pi/l_{osc}$, in the momentum
integration, where
$l_{osc} = \rho^{-\frac 1 3}$ is the distance between the oscillators, and a
cutoff in the discrete longitudinal momentum variable at
$N=L/l_{osc}=L \pi p_{max}$.
This means that we have
$(p_{max}/\pi)^3=\rho$, with
$\rho$ as the 3D oscillator density.

\subsection {The low-momentum contribution}
The important low momentum
contribution comes from the term with two photons in the intermediate state.
The energy denominator then vanishes when the momenta of the intermediate
photons match the ones of the external photons. Since all photons then are
low momentum photons, for the leading contribution we can replace
$\omega(p)$ by the low energy expression
$m+p^2/2m$ (and $\omega(k)$ by $m+k^2/2m$) to get the energy
denominator on the form
\be{photon}
\frac{1}{\omega(k_1)+\omega(k_2)-\omega(p_1)-\omega(p_2)}\rightarrow
\frac{2m}{k_1^2+k_2^2-p_1^2-p_2^2}\, .
\ee
Here $p_1$ and $p_2$ are the transverse photon momenta of the intermediate
state. In this case only the lowest longitudinal mode has to be included in the
intermediate state. In the same way as for high momenta there are
corrections, but they are supressed by factors $\eta^2/m^{2}$ or
$\Delta^{2}/m^2$, and we shall neglect them.

We note that an integration of the intermediate state momentum of
this term alone gives rise to a logarithmic ultraviolet divergence.
Eventually this
divergence is of course cut off by the interparticle distance, but before
that the integrand is suppressed by the
factor
\be{approx2}
\sqrt{\frac{\omega(p)}{(\omega(p)^2-\omega_0^2)^2+\eta^2
\omega_0^2}}\approx
\sqrt{\frac{m} {(2m\Delta-p^2)^2+ \eta^2 \omega_0^2} }
\ee
which will introduce an effective cutoff in the momentum integral at
$2m|\Delta|$ and thus provide a scale for the logarithm. If this is
introduced as
an explicit cutoff, the momentum integral, after the angular integration, gets
the form
\be{int2}
   I &=& 2 \pi\int_0^{2m|\Delta|} dp^2
    \frac{1}{(k^2 - p^2 + i\e)}
= -\frac{1}{(2m\Delta)^4} \left( \ln\frac{k^2}{2m|\Delta|}-i\pi\right)
\ee
There will also  be a constant ($k$ independent) contribution, but as shown
explicitly in Appendix B this is generally small compared to the leading high
momentum contribution \pref{M2a}.
With the relevant constants and symmetry factors included, the logarithmic
contribution to ${\cal T}$ is, for red and blue detuning,
\be{M2c}
{\cal T}_{2c}(k^2) =  \frac g 4 \left(\ln\frac {k^2}
{2m|\Delta|} -i\pi \right) {\cal T}_1
\ee

There is also a term corresponding to one photon and one dipole in the
intermediate state, but the real part of this is subleading relative 
to the terms
already included and can therefore be omitted. However, for blue
detuning it has
an imaginary part
\be{M2b2}
{\cal T}_{2b}^{blue}(k^2) = i\pi g \left(\frac\eta\Delta\right)^2
{\cal T}_1\, .
\ee
Although small compared to the leading contribution, this
is the dominant imaginary part corresponding to the decay process
$\gamma\gamma\rightarrow\gamma D$ discussed earlier.
As a check on our calculations, we have verified that this imaginary part of
the  scattering amplitude is correctly related to the total inelastic cross
section
in \pref{cross}.

\subsection {The  scattering amplitude  }

Combining \pref{M1}, \pref{M2a} and \pref{M2c} we get the following
approximation for the (one loop) scattering amplitude corresponding
to red detuning
\be{scattamp}
{\cal T}(k^2) =
   \frac {g}{2\pi m} \left[ 1 + \alpha +
\frac g 4 \left(\ln\frac {k^2} {2m\Delta} -i\pi \right)\right] \, .
\ee
The expression is consistent with the expression for the scattering
amplitude of a renormalized delta function interaction, when expanded to
second order in the coupling strength. Resumming the diagram 2a as a
geometrical
series in fact gives the full scattering amplitude
\pref{sca}, if we set $g_r=g$ and define the renormalization scale
$\mu_r=\mu_0$
by
\be{rsc}
\mu_0^2 = 2m|\Delta|e^{-\frac {4\alpha} g }\, .
\ee
The same expression
is valid for blue detuning if we neglect the effect of scattering
$\gamma\gamma\rightarrow\gamma D$. Note, however, that the
sign of $\alpha$ is different in the two cases. The value of the exponent is
\be{exp}
-\frac {4\alpha} g =\half(2\pi)^3\frac{\rho \Delta^3}{m^2\eta^4}
\ee
and we note that this (in absolute value) will be much larger than $1$ with
the assumptions about parameter values which we have made.

The choice of renormalization scale, $\mu_{0}$ in \pref{rsc} is
however misleading in that the logarithm becomes large.  The definition
$\mu_r = 2m|\Delta|$ amounts to  a more natural
choice since  the interpretation of the photons as massive non-relativistic
particles only makes
sense for $k^2/(2m) < |\Delta|$.
The corresponding value for the renormalized coupling constant is
\be{coup}
g_r=\frac{g}{1-\alpha},
\ee
and relative to the bare (first order)
coupling constant $g$, the change in $g_r$ remains small as long 
as the logarithm is small
and $\alpha \approx\Lambda/(m^2\Delta) << 1$. But
depending on the parameter values, $\alpha$ may in reality become 
large and give
rise to a significant renormalization effect. For large $\alpha$ we have
\be{gr}
g_r \approx -\frac{g}{\alpha} 
= \frac{1}{\pi^3}\frac{m^2\eta^4}{\rho\Delta^3}
\ee
and we note that in this limit the effective interaction is independent
both of momentum and of
the anharmonicity parameter $\Lambda$ of the oscillator spectrum. The
detuning parameter $\Delta$ (and not the anharmonicity parameter $\Lambda$) now
determines the sign of the effective coupling, with repulsive
interaction for red
detuning and attractive interaction for blue detuning.


\subsection{Comparison with the Kerr nonlinear susceptibility
coefficient $\chi^{(3)}$ for two-level atoms}

Since the renormalization parameter $\alpha$ is proportional to $1/\Delta$, it
easily gets large for small detuning, $\Delta<<\omega_0$, as shown explicitly
for the case of photons interacting with Rydberg atoms, in the section below.
The renormalized coupling constant $g_r$ (which then is much smaller than the
bare coupling constant $g$) should then be interpreted as the physical
interaction parameter. One should, however, note that the expression we
have found for $g_r$ is not based on a systematic expansion in $\Lambda$,
but rather by resumming ``dangerous" terms in the expansion. There will be
other contributions to the renormalized coupling, but these
are parametrically small, \ie supressed by powers of small
ratios like $\eta/m$ or $\Delta/m$. Without resumming other parts of
the perturbation series, we cannot determine in which parameter range
these terms can be neglected. For the following estimates we shall
simply assume that we are in that range.

The interpretation of $g_r$ as the physical interaction parameter is
reinforced by the fact that the expression we have found
\pref{gr}, depends on the detuning parameter $\Delta$,
the effective plasma frequency
$\eta$, and the atomic number density $\rho$
in exactly the same way as the non-linear Kerr
susceptibility coefficient $\chi_{G}^{(3)}$ for a dilute gas of
two-level atoms, as obtained  by Grischkowsky \cite{Grischkowsky},
\begin{equation}
\chi_{G}^{(3)}=\frac{\rho\mu_e^{4}}
{2\Delta^{3}}\label{Grischkowsky}
\end{equation}
with $\mu_e$ as the dipole matrix element (for one component of the
dipole vector)
connecting the two states of the two-level atom. Rewritten in our notation,
\begin{equation}
\mbox{$\label{mu}$}
\mu_e^{2}=\frac{q^{2}}{M\omega_{0}}=
\frac {\pi^{3}} 2 \frac {g_{r}}{m^{2}\omega_{0}^{2}}
\end{equation}
which gives
\begin{equation}
\mbox{$\label{Gris2}$} \chi_{G}^{(3)}=\frac{\eta^{4}}{2\rho\omega_{0}
^{2}\Delta^{3}}=\frac{\pi^{3}}{2\omega_{0}^{2}}g_{r}\text{ .}
\end{equation}
To compare the expressions, we write the 3D
susceptibility, extracted from our effective action \pref{eff4b}, in terms of
the dimensionless bare coupling constant $g$,
\begin{equation}
\chi^{(3)}=\frac{1}{16}\frac{\eta^{4}\Lambda}{\rho m^{4}\Delta^{4}
}=\frac{(2\pi)^{2}}{3}\frac{g}{m^{4}}\label{sus} \, .
\end{equation}
If we assume that the 3D susceptibility renormalizes in the same way as the
2D dimensionless $g$ (the renormalization comes from high $k$ where the
dimensional reduction from 3D to 2D is not relevant), then
\be{kerr}
\chi_{r}^{(3)}=\frac{(2\pi)^{2}}{3}\frac{g_{r}}{m^{4}}
= \frac 8 {3\pi} \left(\frac {\omega_{0}} m \right)^{2}\chi
_{G}^{(3)} \, .
\ee
and since $m\approx\omega_0$,  our expression for
$\chi ^{(3)}$ is very close to Grischkowsky's.

That the  factor $8/3\pi \approx .85$ is very close to one
is of no significance,  since our coefficient depends on the details of the
ultraviolet cutoff. What is relevant, however, is that the two quite
different approaches give essentially the same result for the
susceptibility, and also that our result is independent of the bare
non-linear coupling.

\section {6  Scales, Bose-Einstein condensation and two photon bound states}

Based on the discussion of the effective photon-photon
interaction in the previous sections, we will now consider some of the
physical aspects of the formation of a two-dimensional photon fluid. We first
summarize the important parameters that characterize the photon
system, and then
discuss the conditions under which two particularly interesting
phenomena could
occur: the formation of a two dimensional  Bose-Einstein condensate, and the
formation of two-photon bound states. We should  stress that both these effects
are essentially quantum mechanical, and cannot be described  by classical
non-linear optics.

\subsection {Important scales}

The strength of the mixing between the photons and the oscillators is
given by the effective plasma frequency $\eta$ defined by
\be{etadef}
\eta^{2} = \frac {\hbar c q^2\rho} M \, ,
\ee
and mixing becomes important when $\eta^2 \sim \Delta^2$.
From now on we shall restore factors of $c$ and $\hbar$ in the
formulas.

The (unrenormalized) interaction strength is given by
the  dimensionless coupling constant $g$,
\be{gdeff}
g = \frac 1 {(2\pi)^2} \frac 3 {16} \frac {\hbar\Lambda }{c^{3}\rho}
\left(\frac {\eta} {\Delta}\right)^4 \, .
\ee
The importance of the non-linear loop corrections to the interaction strength
is, for momenta $k^2\sim 2\hbar m\Delta$, determined by the dimensionless
parameter
$\alpha$,
\be{alph}
\alpha =   -\frac {3\pi} {64} \frac {\hbar^{3} \Lambda} {c^{4}m^2\Delta}   \, .
\ee
and for large $\alpha$ the effective coupling constant is given by
\be{grny}
g_r \approx \frac{1}{\pi^3} \frac{cm^2\eta^4}{\hbar^{2}\rho\Delta^3} \, .
\ee
Finally,  for very small momenta, with
\be{kcond}
g\ln\frac {k^2} {2m\Delta} \sim  1 + \alpha \, ,
\ee
the logarithmic term will become important and contribute to the
renormalization of the interaction.

\subsection {Photonic Bose-Einstein condensates}
As mentioned in the introduction, one of the most exciting
aspects of forming a photon gas, would be the possibility
to study phase transitions. First, recall that there is no Bose-Einstein
condensation (BEC) in a free two dimensional gas. However, the
situation is different when the gas is in a trapping potential,
where the transition temperature is given by $k_{B}T_{c}\approx
\sqrt{N}\hbar \Omega$
\cite{Bagnato91}, where $N$ is the total number of bosons and $\Omega$ is the
frequency of the (harmonic) trapping potential.  In an interacting gas the
situation is more complicated. The Mermin-Wagner theorem
\cite{Mermin66} rules out
true long-range order, but  there is still the possibility of a
Kosterlitz-Thouless (KT) transition, with the formation of a
``quasicondensate''
at a critical temperature
$k_{B}T_c \approx \rho_{2D} \hbar^2/{m} $, with $\rho_{2D}$  the 2D boson
density and $m$  the boson mass. In the context of two-dimensional
atomic BEC,
the phase transition in a quasi two dimensional Bose gas with repulsive delta
function interactions
was recently analyzed by Petrov, Holzmann and Shlyapnikov
\cite{petrov}. They find that in a trapping potential a true condensate will
form well below $T_c$, whereas for intermediate temperatures $T<T_c$ it will
change to a quasicondensate with a spatially fluctuating phase.

For a two-dimensional (dilute) photon gas a similar analysis should
be relevant. We will make some comments on this in the discussion
of a photon gas interacting with Rydberg atoms in the next section.
For this discussion the expressions for the effective interaction
parameter, \pref{grny} and the two-dimensional cross-section \pref{cross} are
important. There are also important questions concerning the
thermalization time and the possibility of regulating the effective
temperature of the photon gas. With the parameter values discussed below
the critical temperature is typically very much higher than the excitation
energies. This indicates that a thermalized photon fluid would tend to form
a Bose condensate rather than a normal fluid, and that the phase transition
temperature cannot easily be reached by varying the parameter values of the
photon fluid.

\subsection {A two-dimensional two photon bound state}

The scattering amplitude
\pref{sca} always has a pole for complex $k$ on the second
Riemann sheet, at $k^2 = -2mE_b$.
This corresponds to a bound state, a ``diphoton state" \cite{diphoton}, with
binding energy $E_b$ given by
\be{bs}
E_b = \frac {\mu_0^2} {2m} e^{\frac 4 g} = |\Delta|
     e^{\frac {4} {g_r} }  \, .
\ee
Somewhat surprisingly it looks like one could have a bound state
for either sign of $g_r$, and this is indeed the case for a
fundamental delta function interaction \cite{jackiw1}. We note,
however, that for repulsive interactions ($g_r>0$), the bound state
would occur at large $ k$, far outside the range of validity
of our effective theory.
For attractive interaction ($g_r<0$) however,
the presence of such a two photon bound state is a bona fide
quantum mechanical effect with no obvious counterpart as a solution
of the non-linear Schr{\"o}dinger equation.

When $|\alpha|<1$, attractive interaction corresponds to
negative $\Lambda$, and diphoton states could in principle
form for red detuning, where the photons are stable against
decay to the lower energy dipole mode. However, when
$\alpha>>1$, which seems most relevant for physical realizations (se
discussion below), weakly bound photon pairs would occur only for
blue detuning. In this case we do not expect the formation of stable
bound pairs due to the presence of the decay channel to the dipole
mode.

\section{7  The 2D photon fluid: Physical realizations}

In this section we shall discuss two possible experimental
scenarios for the formation of a 2D photon fluid.
First, we  consider microwave photons in a cavity
filled with Rydberg atoms. This scenario is quite close to the one described
in the previous sections, since the highly excited atoms are well described
as harmonic oscillators with a small anharmonicity. As a second example we
shall consider optical photons in a cavity filled with alkali atoms in their
ground states.

The question we shall address is the following: Can the parameters of
these systems be tuned in such a way that there results a sufficiently large
effective photon-photon interaction---mediated by virtual transitions within
the atoms---so that a two-dimensional photon fluid can form within a
Fabry-Perot resonator? One might expect that such a fluid arises after many
effective photon-photon collisions within the cavity.
Based on the analysis of the previous sections we reach an answer
to this question which is qualified affirmative. The conditions for a
photon fluid in thermal equilibrium to form may be met, but only under
conditions where the parameters can be tuned to optimal values. For the
Rydberg atoms these conditions seem more difficult to satisfy than
for the optical
photons.

However, the conclusions concerning the formation of a photon fluid
are based on
simple order of magnitude estimates of the physical parameters. One should note
that the expressions used in these estimates were found above
assuming certain
conditions (separation of physical scales) which are not necessarily met in the
real system. Also the expression found for the interaction parameter
$g_r$ is based on the resummation of certain large contributions, while other
higher order contributions are left out. For this reason we
can present only a preliminary
evaluation of the conditions for a 2D
photon fluid to form. A detailed evaluation of these conditions is outside the
scope of the present paper.

\subsection{Rydberg atoms in a microwave Fabry-Perot cavity: Numerical
estimates}

We first consider the case where microwave photons interact
via Rydberg atoms excited near resonance, with both photons and atoms
injected into a microwave version of a Fabry-Perot cavity. Consider a
ribbon-shaped  beam of Rydberg atoms passing through the gap between a
pair of parallel, conducting plates, separated by a spacing $L_{L}
\approx\lambda_{0}/2$. The plates are partially transparent to
microwaves,
for example, by having a regular square array of small holes, and
 can thus act as  the (highly reflective) mirrors of a microwave
Fabry-Perot cavity. Microwave photons can  be injected into the
cavity from the
outside, and their transverse distribution, after many photon-photon
interactions mediated by the atoms, can be monitored in transmission. The
resonance frequency of the cavity can be tuned to  near a resonance
frequency of the Rydberg atoms.

For simplicity, let the plates be square in shape, with a typical transverse
dimension of $L_{T}>>\lambda_{0}$, where $\lambda_{0}$ is the microwave
resonance wavelength of the Rydberg atom transition of interest. The
longitudinal mode number of the cavity is restricted to that of the
fundamental longitudinal mode of the cavity (by our choice of spacing
$L_L\equiv L=\lambda_{0}/2$ between the plates), but there can exist
many possible
transverse modes of the cavity, which are closely spaced near the fundamental
longitudinal mode.

For concreteness, we shall use as a guide the parameters of the experiment of
the Paris group \cite{Gawlik}, where Rubidium
atoms were excited to $n=50$, where
$n$ is the principal quantum number of the Rydberg atom. The Rydberg atoms can
be put into the ``circular'' state $|n,l=n-1,m=n-1\rangle$, which have a
maximal electric dipole moment, and thus a maximal coupling to the microwave
photons. Since the typical size of such a Rydberg atom is given by the radii
of the circular Bohr orbits $a_{n}=n^{2}a_{0}$, where $a_{0}\approx0.5$
\AA\ is the Bohr radius, the atom can be quite large in size, e.g.,
$a_{n=50}\approx0.125$ microns. The electric dipole transition matrix element
for the $n=50\rightarrow n=51$ transition is also quite large, and therefore
the photon-photon coupling mediated by such atoms virtually excited near
resonance should be correspondingly large.

In the regime of high $n$, the energy levels of Rydberg atoms are almost
equally spaced,
\begin{equation}
E_{n+\Delta
n}=-\frac{\text{Ry}}{(n+\Delta n)^{2}}=E_{n}+\frac{2\text{Ry}}{n^{3}}
\Delta n-\frac{3\text{Ry}}{n^{4}}\Delta n^{2}+\dots\label{sp}
\end{equation}
where $Ry = 13.6$eV is the Rydberg constant. The dominant photon-induced
transitions
will be between the circular states,
$|n,n-1,n-1\rangle\rightarrow|n\pm1,n\pm1-1,n\pm1-1\rangle$,  where the wave
function of the state $|n,n-1,n-1\rangle$ is
\[
\mbox{$\label{ryc}$}  \psi_{n}(r,\theta,\psi)=\left[  \frac{(2\kappa)^{3}
}{8\pi n}\right]  ^{\frac{1}{2}}\frac{(-1)^{n-1}}{2^{n-1}(n-1)!}e^{-\kappa
r}\,(2\kappa r)^{n-1}\,(\sin\theta)^{n-1}\,e^{i(n-1)\phi}\,,
\]
with $\kappa=1/na_{0}$.
This system is very well
suited to be modeled by an anharmonic oscillator model of the type discussed
earlier. We shall determine the anharmonic parameter
$\Lambda$, the oscillator frequency $\omega$, and the oscillator mass
$M$ of this
model in terms of the Rydberg atom parameters, by matching the both
the energy spectra  to terms quadratic in $n$ and the dipole matrix
elements between the levels $n$ and $n+1$.

One might object to this procedure since in our previous calculations we
assumed the oscillators to be in their ground state rather than in a highly
excited state. In particular this means that the effect of virtual transitions
to lower Rydberg states that would be present in Rydberg atoms with the
spectrum\pref{sp} are not taken into account. However, it should be kept in
mind that the detuning for transitions of Rydberg atoms to lower states will
be substantially larger than those to the higher states. In any case, for the
simple numerical estimates made in this section we shall ignore this effect,
and assume that only upward transitions $n\rightarrow n+1$ are important. This
approximation will at most give rise to a numerical factor of order of unity in
the estimate of the model parameters. For the same reason we shall neglect
the fact that the oscillators of our model are three-dimensional and
simply match the dipole matrix elements of the Rydberg states with those of the
corresponding one-dimensional anharmonic oscillator.

With the above matching procedure for ``circular'' Rydberg atoms prepared in a
state with the initial principal quantum number $n$, we get the following
result for the anharmonicity parameter $\Lambda$ of the oscillator model
\begin{equation}
\Lambda=-\frac{8\omega_{0}^{2}\text{Ry}}{\hbar^{2}n^{4}}
=-\frac{4\omega_0^3}{n\hbar}\text{ .}
\end{equation}
However, this expression gives for the
magnitude of the renormalization constant,
$\left|  \alpha\right|  \simeq$ $\omega_{0}/(n \Delta)$, and $\alpha$
therefore becomes very large, since we shall operate the system with very small
detunings near resonance, so that
$\Delta<<\omega_{0}$. The detuning $\Delta$, however, must be much
larger than the
natural line width of the atoms \cite{Einstein-A}, and in this limit we
can use the $\Lambda$ independent expression \pref{grny} for the
renormalized coupling constant.
Note that in this limit the sign of the physical
interaction is determined solely by the detuning parameter
$\Delta$ - repulsive for red detunings $\Delta>0$, attractive for blue
detunings, $\Delta<0$.
Matching the dipole matrix elements effectively amounts to introducing a
large oscillator strength, 
$f_{osc}=n^2(2n-1)/(n-1)^2$  
in the expression for the effective plasma
frequency $\eta=(q^{2}\rho/M)^{1/2} \rightarrow
(\hbar c  e^{2} f_{osc}\rho /m_{e})^{1/2}=c(\alpha_{e}f_{osc}\lambdabar_{C}\rho
_{atom})^{1/2}$, where $\alpha_{e}$ is the fine structure constant, $m_{e}$
is the
electron mass, and $\lambda_{C}=2 \pi \lambdabar_C$ is the Compton
wavelength of
the electron.
Although, strictly speaking, the detuning $\Delta$ should be much greater than
the effective plasma frequency $\eta$, for the purposes of numerical
estimates, we shall set $\Delta$ equal to $\eta$. Also, we shall approximate
$m c^2\approx\hbar \omega_{0}$, where $\omega_{0}$ is the atomic resonance
frequency of the transition in question. With this we get,
\begin{equation}
g_{r}=\left(  \frac{2}{\pi}\right)  ^{3/2}\left(  \frac{\alpha_{e}f_{osc}
}{\rho_{atom}\lambda_{0}^{3}}\frac{\lambda_{C}}{\lambda_{0}}\right)
^{1/2}\text{ ,}\label{g_r(fine-structure)}
\end{equation}
where $\lambda_{0}$ is the atomic resonance wavelength of the
transition in question.

As a first estimate of the size of $g_{r}$, consider the case of the the
$n=50\rightarrow n+1=51$ ``circular'' Rydberg transition, where
$mc^2/\hbar\approx\omega_{0}=2\pi\times51$ GHz, with $\rho\simeq1000$
atoms per $cm^{3}$ and $f_{osc}=105$. This yields an effective plasma
frequency of $\eta\simeq 5.15\times 10^{6} s^{-1}$. 
For this transition, assuming that
$\Delta\simeq\eta$, we find that $g_{r}\simeq 6.32\times10^{-7}$.

\subsection{Cavity dimensions }

To fit the frequency of the Rydberg transition $n=50\longrightarrow51$,
i.e., $\omega_{0}=51$ GHz, we assume the dimensions of the cavity to be
chosen as follows,
\begin{equation}
L_{L}=\frac{1}{2}\lambda_{0}\approx3\text{ mm and}\;L_{T}\approx100\text{
cm,}\label{LL}
\end{equation}
with $L_L=L$ as the longitudinal extension of the cavity and $L_T$ as the
extension in the transverse directions. The limits for the transverse momentum
$k_T=k$ are therefore
\begin{equation}
\frac{\pi}{L_{T}}<k_{T}<\left(  2m\Delta\right)  ^{1/2}.\label{limits}%
\end{equation}
For a detuning $\Delta\simeq 2\pi\times2$ MHz, this implies that
there are around
10 nonrelativistic transverse modes in the cavity which can interact via the
Rydberg atoms. Higher modes will be available and they will also
interact via the atoms, but the energy-momentum relation for these higher
modes will be modified relative to the non-relativistic expression. However,
even with the limited number of modes available one should in principle be able
to see the formation of a photon fluid in a Bose condensed state.

\subsection{The effective photon-photon collition rate and
thermalization}

We showed earlier that the total elastic cross-section in 2D is given
by
\begin{equation}
\sigma^{el}\approx\frac{g_{r}^{2}}{k_{T}}\text{{.}} \label{cross2}
\end{equation}
 The ``reaction rate'' for photon-photon collisions in 2D is
therefore
\begin{equation}
\sigma^{el}v_{T}=\sigma^{el}\frac{\hbar k_{T}}{m}=\frac{\hbar}{m}g_{r}
^{2}\text{ .}\label{rate}
\end{equation}
The
collision frequency $\omega_{coll}$ for photon-photon collisions in 2D is
\begin{equation}
\omega_{coll}=\rho_{2D} \sigma^{el} v_{T}=\rho_{3D}L_{L}\sigma^{el}v_{T}\text{
,}\label{w_coll}
\end{equation}
where
$\rho_{2D}$ is the 2D photon number density, and
$\rho_{3D}\,$is the 3D photon number density. Now the number of photon-photon
collisions $N$ that occur within a cavity ring-down time $\tau$
is given by $N=\omega_{coll}\tau$, where, given the quality
factor $Q$ of the cavity, $\tau=Q/\omega_{0}$. Solving for
the number density of photons $\rho_{3D}$ needed for $N_{coll}$ photon-photon
collisions, one obtains
\be{density}
\rho_{2D}&=&\frac{N}{Q    \,g_r^2}{\left(\frac{\omega_0}{c}\right)}^2=
4\pi^2\,\frac{N }{Q    \,g_r^2}\,\lambda_0^{-2} \nn
\rho_{3D}&=&8\pi^2\,\frac{N }{Q    \,g_r^2}\,\lambda_0^{-3} \, .
\ee
We note that due to the smallness of the interaction parameter
($g_{r}\simeq 6.3\times10^{-7}$), there is a large numerical factor in these
expressions. If we assume
$N \simeq10$ and  a rather high quality factor of $Q    \simeq10^{7}$,
\be{numfac}
\frac{N }{Q    \,g_r^2}=2.5\times 10^{6}
\ee
and the corresponding value of the photon density therefore gives a
KT transition temperature far above the low-momentum regime of the
interacting photons,
\be{KTT}
k_BT_{c}\approx\frac{\rho_{2D}\hbar^2}{m}\approx 2\times10^6
\,\hbar
\omega_0
\ee
corresponding to a temperature $T_c\simeq 4\times10^6$ K. Thus, the low-energy
photons are typically in a temperature interval where a Bose
condensate will form,
rather than a normal fluid.

The microwave intensity inside the cavity is
\be{intensity}
I_{in}=\rho_{3D}\,c\,\hbar\omega_{0}
\ee
and with the given parameter values we find $I_{3D}\approx 1.1$ mW cm$^{-2}$.
To convert from the inside-cavity to the outside-cavity intensity, we
use the relation $I_{out}=I_{in}/Q    $.
Note that the incident power $I_{out}$ needed in order to form the photon
fluid scales inversely as the $square$ of the quality factor, i.e.,
$I_{out}\sim Q    ^{-2}$, since $I_{in}$ is inversely
proportional to $Q    $.

Based on the above, we estimate that about
1.1 microwatts of incident microwave power is needed
in order to get around ten
photon-photon collisions within a cavity ring-down time of  30
microseconds. Under such circumstances, we expect that a 2D photon fluid
should form inside the cavity.

The value of $Q    \simeq10^{7}$ is feasible ($Q    \simeq10^{8}$ by
means of superconducting cavities have already been achieved \cite{Haroche}).
Note, however, that if $\Delta>\eta$ the situation rapidly deteriorates due to
the dependence on the factor $\left(  \eta/\Delta\right)  ^{6}$ of $g_r^2$.

\subsection{Chemical potential and speed of sound of the photon fluid}

From the previous section it might seem that it is always possible to
form a photon fluid just by increasing the intensity of the incident
microwaves. This conclusion is however not justified, since a too high
intensity will make the fluid
so dense, and  the mean kinetic energy is so large, that the
non-relativistic approximations are no longer valid. To get a
rough estimate of when this happens, we  use an
effective Ginzburg-Landau (GL) type theory that should be applicable
when the fluid is very dense. The relevant potential is given by
\be{thf}
V=-\mu\left|  \phi\right|  ^{2}+\pi g_{r}\frac{\hbar^{2}}{2m}\left|
\phi\right|  ^{4}\approx0\text{ .}\label{V(phi)}
\ee
where we introduced a chemical potential which, in the Thomas-Fermi
approximation, is related to the photon density by  the relation
\be{cpot}
\mu=\pi g_{r}\frac{\hbar^{2}}{2m}\rho_{2D} \, ,
\ee
which is obtained by minimizing $V$ with respect to
$|  \phi | ^{2}=\rho_{2D}$. The GL theory defined by $V$
implies sound waves in the photon fluid with a speed given by
\cite{Chiao2000},
\be{ss}
v_{s}=\left(  \frac{\mu}{m}\right)  ^{1/2}\text{ .}\label{v_s}
\ee
We can also estimate the mean velocity of the photons using
the virial theorem,
\be{vir}
\langle E_{kin} \rangle = \frac {mv_{\gamma}^{2} } 2 \rho_{2D} =
\langle E_{pot} \rangle = \mu \rho_{2D}
\ee
which gives $v_{gamma} = \sqrt{2} v_{s}$.

Imposing the consistency condition that
$v_{s}=(\mu/m)^{1/2}<c$, we obtain an upper limit
on the 2D photon number density that $\rho_{2D}<2m^{2}c^{2}/\hbar^{2}\pi
g_{r}$. Combining this with the formation condition that $N >>1$, and
using $\omega_{0}=mc^{2}/\hbar$, we obtain the upper and lower bounds
on
$\rho_{2D}$
\begin{equation}
\frac{m^{2}c^{2}}{\hbar^{2}Q    g_{r}^{2}}<<\rho_{2D}<\frac{2m^{2}c^{2}
}{\hbar^{2}\pi g_{r}}\text{ .}\label{upper-and-lower-bounds}
\end{equation}
We therefore conclude that a self-consistency condition for the formation of
the photon fluid is
\begin{equation}
Q    >>\frac{\pi}{2}g_{r}^{-1}\text{ .}\label{self-consistency}
\end{equation}Thus, the choice of a high-$Q$ resonator is an
important criterion
for the formation of a photon fluid. For the case of the Rydberg transition
$n=50\rightarrow51$, we see that it would be necessary for $Q    >>10^{6}$,
so that the choice of
$Q
\simeq10^{7}$ may be sufficient.

\subsection{Alkali atoms in an optical Fabry-Perot cavity}

The apparent advantages of using microwaves
and Rydberg atoms, rather than
a more standard setup with lasers and \eg Rubidium atoms,
is twofold. First the cavity dimensions are not too
small, and second, the large dipole moment of the atoms implies a
large nonlinearity. The obvious disadvantage is that the Rydberg atoms
are much harder to create and manipulate. Also since we have shown
that
the anharmonic parameter $\Lambda$ drops out of the renormalized
coupling constant $g_r$ for small $\Delta$, it is of interest to
consider an optical Fabry-Perot cavity  with
alkali atoms in their ground states, for which there is only a single
transition to
one excited state with a very strong oscillator strength of the order of
unity, which almost completely exhausts the $f$-oscillator sum rule. For alkali
atoms, a two-level model for the atom is therefore a more appropriate one
\cite{Grischkowsky}. However, the renormalized coupling constant for
photon-photon interactions, and the resulting rates for collisions found in
the previous section, should still apply here, provided that we remember to use
the condition that the detuning $\Delta$ be comparable to the effective plasma
frequency $\eta$ for the expressions for collision rates.

Consider the case of a fundamental longitudinal mode Fabry-Perot cavity with
$L_{L}\approx\lambda_{0}/2$. In this case, the $Q    $ of the cavity is also
approximately its finesse. High quality mirrors with
$Q    \simeq10^{5}$ are commercially available \cite{PMS}. For a
strongly allowed
alkali transition,
$f_{osc}\simeq1$. If we assume a density $\rho_{atom}\simeq10^{9}$
rubidium atoms
per cm$^{3}$ (rubidium atoms have a transition wavelength of $\lambda_{0}=780$
nm), this leads to an effective plasma frequency $\eta\simeq 2\pi\times 200$
MHz (which is comparable to the Doppler width of rubidium atoms at room
temperature), With $\Delta\simeq\eta$ this gives $g_{r}%
\simeq 12\times10^{-3}$. The minimum required $Q    $ is therefore around
130, which is easily satisfied. Thus, if we set $N \simeq10$ and
$Q    \simeq10^{5}$, from Eq.(\ref{intensity}) we obtain $I_{3D}\simeq10^{6}$
W cm$^{-2}$, or an outside power requirement of a 0.1 Watt per square
millimeter
incident on the Fabry-Perot, which is feasible.

One consequence of going from microwave to visible wavelengths, is that the
number of nonrelativistic transverse modes given by Eq.(\ref{limits}) can be
much larger. For the given parameter values
one finds
that there are around 200 nonrelativistic transverse modes which can be
coupled together. Also the estimated KT transition temperature is relatively
lower, with $k_BT_{c}\approx0.7\,\hbar\omega_0$.

\subsection{High-$Q$ microspheres of glass and the formation of a 2D
photon fluid}

Let us finally comment on a third possible way in which a 2D photon fluid could
form. Extremely high-$Q$ optical cavities with
$Q    \simeq10^{10}$ have been fabricated out of low-loss,
small-diameter glass
microspheres \cite{Ilchenko}. A 120-$\mu$m diameter glass microsphere
immersed in
superfluid helium has been observed to exhibit a nonlinear, dispersive bistable
behavior with a threshold power of as low as 10 microwatts, due to
the intrinsic
Kerr nonlinearity of the glass which constitutes the microsphere
\cite{Haroche1998}. Due to the curvature of the microspheres the photons are
tightly confined to propagate within an optical wavelength or so of the
two-dimensional spherical surface, \ie within its
``whispering-gallery'' modes. This tight, two-dimensional confinement
of the light
should also allow an effective dimensional reduction from 3D to 2D in
the photon
degrees of freedom, which differs from the mechanism of dimensional
reduction in a
planar Fabry-Perot cavity discussed here in this paper. Thus a two-dimensional
photon fluid could also in principle form  just within the inside
surface of the
microsphere, due to the photon-photon scatterings mediated by the
intrinsic Kerr
nonlinearity of the glass, and may have already been indirectly observed in the
experiments of
\cite{Haroche1998}.
It should be noted that
the change of topology from that of a plane to that of a sphere should not make
the formation of the 2D photon fluid impossible in principle.

\section{8  Concluding remarks}

In this paper we used a microscopic approach to study the effective
interaction between photons confined to a quasi-two-dimensional
cavity filled with
a non-linear medium. With the atoms modelled by a collection of non-linear
Lorentz oscillators, the interaction was studied to second order in
perturbation theory. The linear problem was first solved by decoupling the
``polariton modes", and we  described efficient ways to do this,
both by use
of a path integral approach and by use of exact diagonalization, as shown in
Appendix A. A main motivation was to examine the description of
the effective
interaction, induced by the non-linearity, as a 2D short range (delta function)
interaction and to determine the renormalization scale associated with this
interaction. We  noticed several interesting complications. For
red detuning
relative to the oscillator frequency, the expression found to second order fits
with the form of a normalized delta-function potential. However, the
renormalization parameter is typically large and a resummation of ``dangerous"
terms has been performed in order to extract an effective, renormalized
interaction strength. This way of including higher order contributions raises
questions of whether other contributions, not included here (since
they are parametrically small), may also be of importance. However,
additional support for the expression found for the renormalized interaction
parameter comes from other evaluations of the non-linear
susceptibility of a gas of
alkali atoms. For blue detuning, we discussed the possible formation
of photonic bound states, and noticed the possibility that photons
might ``decay"  into oscillator excitations.

The last part of the paper was devoted to a discussion of possible
experimental scenarios, where a photon fluid may form in a
Fabry-Perot resonator.
In these cases, two-dimensionality  is obtained by constraining the
photons in one
direction to their fundamental mode. By means of order of magnitude
estimates we  examined the constraints on the physical parameters in order
for a photon fluid to form. There is a potential conflict between the
need to have
a high density of photons in order to obtain a satisfactory collision frequency
and the possibility of having a too high density measured in units of
the photon
Compton wave length, since the effective photon mass is typically
very small. For
photons in a microwave cavity interacting with Rydberg atoms the corresponding
constraints are serious, although by choosing optimal parameter values the
conditions for creating a 2D photon fluid can probably be met. Such an
experimental realization is otherwise attractive, as the model of photons
interacting with Rydberg atoms is closely related to the oscillator
model used in
the paper. For photons in an optical cavity, the constraints on the physical
parameters seem less severe, although the use of such small scale cavities may
pose a larger experimental challenge. We  noticed that in both cases the
typical temperature associated with a Kosterlitz-Thouless phase transition lies
well above the energy scale associated with non-relativistic 2D
photons, and for
such low energies a condensate will typically form if the collision
frequency is
sufficiently large on the scale set by the ringdown time of the cavity. In
addition to the two possible experimental realizations discussed in
some detail, we  also suggest that other types of realizations of the 2D
photon fluid may
be possible, and  in particular we pointed at the use of high-$Q$
microspheres of glass as being potentially interesting in this context.

Let us finally stress the point that in this study we have made some
simplifying assumptions, in particular about a clear separation of the physical
(energy) scales associated with the formation of a 2D photon fluid. In
real systems,
these conditions may not be well satisfied, and in the discussion of
microwave photons
interacting with Rydberg atoms we  noticed that one may have to tune the
frequency so close to the resonance value that the photons become strongly
mixed with the oscillator degrees of freedom, \ie they are genuine polaritons
rather than photons. For this reason we consider the order of magnitude
estimates applied to these realizations only as preliminary ones. A more
detailed  study is needed to settle more firmly the conditions for the
creation of 2D photon fluids in such small Fabry-Perot cavities.

\subsection{Acknowledgements}
We thank John Garrison and Colin McCormick for helpful discussions. RYC
acknowledges the partial support of this work by NSF grant PHY-0101501. 
JML thanks the Miller Institute for
Basic Research in Science for financial support and hospitality.
Supports from the Research Council of Norway and the
US-Norway Fulbright Foundation are also acknowledged.
THH thanks Nick Khuri and the theory group at Rockefeller University, as
well as Shivaji Sondhi and the condensed matter group at Princeton
University for hospitality.

\section {Appendix A . Alternative derivation of the effective photon action}
\noindent

In this Appendix we derive the effective action \pref{eff2b} and
\pref{eff4b} by an alternative method which does not require
any field transformation of the type \pref{transf} which is non local
in time. In addition to verifying the soundness of the simple procedure
used in the text, the present derivation explicitly shows that the
effective action is the sum of the two contributions in \pref{eff2b} and
\pref{eff4b} corresponding to the two frequencies $\omega_{\pm}$.

The most natural object  to consider given the action \pref{act},
would be the generating functional $Z[\vec J_{\vec A}, \vec J_{\vec
R}]$, which can be written as a Feynman path integral over the
fields $\vec A$ and $\vec R$.
By a change of variables one can diagonalize
the quadratic part of the action, and  then use perturbation
theory to calculate Greens functions. However, becuause of
   the time derivative coupling \pref{dip}, this will give rise to
expressions which are nonlocal in time, since this procedure fails to
correctly describe the normal modes. An obvious way to proceed is to
switch to a Hamiltonian formalism,  do an explicit diagonalization
of the quadratic piece of the Hamiltonian by a symplectic transformation
on the phase space, and finally express the nonlinear term in
the new variables. This method is however cumbersome, and we now present
a simpler approach based on the path integral.

The essential step is to use an alternative form of the action which
is related to
\pref{act} by a Legendre transformation. Rather than just writing down
this transformation, we shall go back to the derivation of the Feynman
path integral, since this will illuminate the physical meaning of the
new variables. The usual Hamiltonian description of \pref{act}
would comprise two canonical pairs $(\vec A, -\vec E)$ and $(\vec R,
\vec P)$, where   $\vec P =M \dot {\vec R} + q\vec A$. Since we are
dealing with a coupled system of matter and radiation, it is rather
natural to introduce the electric displacement vector by
\be{dispxx}
\vec D = \vec E + q\rho \vec R = \nabla\times \vec C \, ,
\ee
where the last identity defines the "displacement vector potential"
$\vec C$ provided that there are no macroscopic charges, \ie
$\nabla \cdot \vec D = 0$. 
The variables $\vec D$ and  $\vec C$ were used earlier by 
Hillery and Mlodinow in the context of quantization of
non-linear electrodynamics \cite{hillery1984}.
It is easy to show that $\vec C$ forms
a canonical pair with the magnetic field strength $\vec B$.
If we furthermore introduce the rescaled oscillator fields
\be{resc}
\vec Q &=& \sqrt{\rho M} \vec R \\
\vec \Pi &=& \frac 1 {\sqrt{\rho M} } \vec R \nonumber \, ,
\ee
the canonical commutation relations take the form,
\be{canbra}
[B_i(\vec r) , C_j(\vec r')]
         &=& i\hbar \delta_{ij}\delta^3(\vec r - \vec r') \\
\left[Q_i(\vec r) , \Pi_j(\vec r')\right]
         &=& i\hbar \delta_{ij}\delta^3(\vec r - \vec r')\, ,  \nonumber
\ee
and the corresponding quadratic Hamiltonian density is,
\be{ham}
{\cal H}^{(2)} = \half (\vec B^2 + \vec D^2 ) + \half \left[\vec \Pi^2 +
(\omega_0^2 + \eta^2) \vec Q^2 \right] -\eta \vec D \cdot \vec Q \, ,
\ee
where again $\vec D = \nabla \times \vec C$, and  $\eta^2 =
q^2\rho/M$.
We now consider $\vec B$ and $\vec \Pi$ as the momenta and derive the
path integral expression for the
generating functional
\be{genfx}
Z[\vec J_D \vec J_R] = \int {\cal D} \vec C {\cal D} \vec Q \,
e^{i\s [\vec C , \vec Q ] + i \int (\vec J_D\cdot \vec D
   + \vec J_R\cdot \vec R )} \, ,
\ee
by first writing a  phase space path integral and then integrating
over the momenta $\vec B$ and $\vec \Pi$. As we shall see below,
the electric displacement field $\vec C$
will in the limit of weak coupling correspond to the photon, and the oscillator
$\vec Q$ to the oscillating dipoles.
At finite coupling these two modes mix and form two
effective propagating "polaritron" modes. In Fourier space, the
quadratic part of the Lagrangian takes the form
\be{klag}
{\cal L}^{(2)} = \half
\left( \matrix{ C(-\vec k)_i &  Q(-\vec k)_i } \right)
\left( \matrix{ P^T_{ij}(\omega^2-\vec k^2) &
i\epsilon_{ijk} \eta k_k \cr
-i\epsilon_{ijk} \eta k_k
    &  P^T_{ij}(\omega^2-\omega_0^2 - \eta^2)   } \right)
\left( \matrix{  C(\vec k)_j \cr  Q(\vec k)_j } \right) \, ,
\ee
where $P^T_{ij}=\delta_{ij} - \hat k_i\hat k_j$ is a transverse projector.
Since in this formulation, the dipole coupling term does not involve any
time derivative, the quadratic part of the action can be diagonalized by
a unitary transformaton which is local in time (although nonlocal in space):
\be{trans}
\left( \matrix{ C_i \cr Q_i } \right) =
\left( \matrix{ \cos\theta_k P^T_{ij} &
             -i\sin\theta_k \epsilon_{ijk}\hat k_k  \cr
i\sin\theta_k \epsilon_{ijk}\hat k_k
& \cos\theta_k P^T_{ij} } \right)
\left( \matrix{  A^{(-)}_j \cr A^{(+)}_j } \right)
\ee
where
\be{angle}
\cos\theta_k = \sqrt { \frac {\omega_0^2 - \omega_-^2}
                                    {\omega_+^2 - \omega_-^2} }
\ee
with $k^2 = \vec k^2$ and, $\omega_\pm (k^2)$ given by \pref{disp2}.

The resulting action is most conveniently expressed in spherical components
defined by $R_m = \hat e_m^* \cdot \vec R$ \etc, with
$\hat e_m^* \cdot \hat e_{m'} = \hat e_{-m} \cdot \hat e_{m'}
=\delta_{mm'}$, $\hat k\cdot \hat e_m =0$ and $m=\pm 1$. We get
\be{final}
{\cal L}^{(2)}(\vec A^{(+)},\,  \vec A^{(-)}) = \half  \sum_{m=\pm 1}
A^{(+)}_{-m}(-\vec k)(\omega^2 - \omega_+^2) A^{(+)}_m(\vec k) +
    A^{(-)}_{-m}(-\vec k)(\omega^2 - \omega_-^2) A^{(-)}_m (\vec k) \, .
\ee
For the interaction term, we must  express  $\vec R$ in terms of $A^{(\pm)}$
using \pref{trans},
\be{rexp}
R_m(\vec k) =  \frac 1 {\sqrt{\rho M}} \left( \cos\theta_k A^{(+)}_m(\vec k)
+ \sin\theta_k A^{(-)}_m(\vec k) \right) \, ,
\ee
where we also redefined $ A^{(-)}_{\pm 1} (\vec k)
                      \rightarrow \pm A^{(-)}_{\pm 1}(\vec k)$.

The final result for the generating function is,
\be{genf2}
Z[\vec J_D , \vec J_R ] = \int {\cal D} \vec A^{(+)} {\cal D} \vec A^{(-)}\,
e^{i\s [A_{a}^{(+)},\,  A_{a}^{(-)}]-
i \int (\vec J_D\cdot \vec D +  \vec J_R\cdot \vec R) } \, .
\ee
with $\s [\vec A^{(+)}, \vec A^{(-)}] = \s^{(2)} +
\s^{(4)}$ explicitly given by
\be{efftot}
{\cal S}_{eff}^{(2)} &=& \frac 1 2  \int \frac{d^{3}k}{(2\pi)^3}\int \frac
{d\omega} {2\pi} \, {\cal L}^{(2)}(\vec A^{(+)},\,  \vec A^{(-)}) \\
{\cal S}_{eff}^{(4)} &=&
         -  \frac {\rho\lambda M^2} {4!}
         \prod_{n=1}^{4} \left[\int \frac{d^{3}k_n\;d\omega_{n}}{(2\pi)^4}
        \right]
   (2\pi)^4 \delta(\sum_{n=1}^{4}\omega_{n}) \delta(\sum_{n=1}^{4}\vec
   k_{n}) \\
       &\times& \vec R(\vec k_1,\omega_{1})\cdot \vec R(\vec k_2,\omega_{2})\,
       \vec R(\vec k_3,\omega_{3})\cdot \vec R(\vec k_4,\omega_{4}) \nonumber
\, ,
\ee
where $\vec R$ is given by \pref{rexp}.
Note that the interaction is non local becauce of the $k^2$
dependence in the denominators
in the $\cos\theta_k$ and $\sin\theta_k$ factors.
Using the identities
\be{rel}
\cos^{2}\theta_k &=& \frac { \eta^{2}\omega_{+}^{2} } {
                                    (\omega_+^2 - \omega_0^2)^{2}
                                    +\eta^{2}\omega_{0}^{2}        } \\
\sin^{2}\theta_k &=& \frac {\eta^{2}\omega_{-}^{2} } {
                                    (\omega_-^2 - \omega_0^2)^{2}
                                    +\eta^{2}\omega_{0}^{2}       }
        \nonumber \, ,
\ee
and substituting the expression \pref{rexp} for $\vec R$
we recover \pref{eff4b}.

\section {Appendix B.
Scattering amplitude to second order}

We perform here a more detailed evaluation of the second order contributions
to the scattering amplitude. The contributions from low and high
momenta of the intermediate states are examined, and we consider in
both cases only the leading contributions in terms of the small
quantities $k/m,
\Delta/m$, and $\eta/\Delta$. The contributions come from diagram a) in Fig.2,
with the intermediate lines corresponding to either the photon or the dipole
mode, with transverse momenta running from zero to the cutoff $1/l_{osc}$,
where $l_{osc}$ is the physical distance between the oscillators. There is also
a sum over all longitudinal momenta, in terms of the discrete momentum variable
$n$, and a sum over the polarization vectors of the fields.

The full expression for the interaction Hamiltonian is
\be{Hintx}
H_{int }=
   K\prod\limits_{r=1}^4 {\left[ {\int {{{d^2k_r} \over {\left( {2\pi }
\right)^2}} }} \right]}\;\delta \left({\sum\limits_r {\vec k_r}}
\right)\;\sum\limits_{\{ n_r\} } {(\delta _{n_1+n_2,n_3+n_4}+2\delta
_{n_1+n_3,n_2+n_4}-4\delta _{n_1+n_2+n_3,n_4})} \nn 
\times \sum\limits_{\{ i_r=\pm\} } \prod\limits_{r=1}^4 [\Lambda_{i_r}\left(
{k_r,n_r} \right) ] {\vec E^{(i_1)}(\vec k_1,n_1,t)\cdot \vec
E^{(i_2)}(\vec k_2,n_2,t)\vec E^{(i_3)}(\vec k_3,n_3,t)\cdot \vec
E^{(i_4)}(\vec k_4,n_4,t)}
\ee
where $K=(2\pi)^2 \Lambda\eta^4/(8\rho L)$ and $\Lambda_{\pm} \left( {k,n}
\right)=[\omega_{\pm}(k,n)^2-\omega_0^2)^2+\eta^2\omega_0^2]^{-1/2}$\,,  $\;n_r
(r=1,..,4$) is the summation variable for the longitudinal mode and $i_r$ for
the photon/dipole mode.
 From this expression the T-matrix and the corresponding scattering
amplitude can be
determined. The T-matrix is related to the scattering matrix by
$S_{fi}=\delta_{fi}-2\pi i\delta(E_f-E_i)  T_{fi}$, where $E_f$ and
$E_i$ are the
energies of the final and initial states, and to second order $T_{fi}$ has
the form
\be{Tx}
T_{fi}=\langle
k'_1k'_2|H_{int}|k_1k_2\rangle+\half \int d^2p_1d^2p_2 \frac{\langle
k'_1k'_2|H_{int}|p_1p_2\rangle\langle
p_1p_2|H_{int}|k_1k_2\rangle}{\omega(k_1)+\omega(k_2)-\omega(p_1)-\omega(p_2)+i\
e}
\ee
where we here have simplified the notation by suppressing 
the sum over field modes, longitudinal momenta
and polarization
variables in the intermediate state.
Galilean invariance implies the following form
\be{Tmatx}
T_{fi}=\delta(K_f-K_i){\cal T}(k_f,k_i)
\ee
where $K_f$ and $K_i$ are the sum of momenta for the outgoing and
incoming particles and
$k_f$ and $k_i$ are the relative momenta. For pure s-wave scattering
(delta function
interaction) ${\cal T}$ only depends on the magnitude of the relative momentum,
\be{calTx}
{\cal T}(k_f,k_i)={\cal T}(k^2)
\ee
and is related to the s-wave scattering amplitude through
\be{sampx}
f=-(2\pi)^2 m\sqrt{\frac 1 {2\pi k}} {\cal T}(k^2)
\ee

For the present case, the first order contribution to ${\cal T}$ is
\be{f1}
{\cal T}_{1}=\frac{g}{2\pi
m}=\frac{3}{16(2\pi)^3}\frac{\Lambda\eta^4}{m\rho\Delta^4}
\ee
With the interaction matrix
elements evaluated and the sum over polarization performed, the
expression for the second
order contribution to ${\cal T}$ is
\be{Tb}
{\cal T} (k^2)&=&
\half C^2\sum_{n_1n_2}\sum_{i_1i_2}(\delta_{n_1n_2}+
\frac{5}{4}\delta_{n_11}\delta_{n_21}+
\delta_{n_1n_2-2}) \nn
&\times & \int
d^2p\left[\frac{\omega_{i_1}(p,n_1)}{((\omega_{i_1}(p,n_1))^2-\omega_0^2)^2+\eta
^2\omega_0^2}
\;\frac{\omega_{i_2}(p,n_2)}{((\omega_{i_2}(p,n_2))^2-\omega_0^2)^2+\eta^2\omega
_0^2}\right.
\nn
&\times&\left. \left(1-\half\frac{p^2}{p^2+n_1^2m_0^2}\right)
\left(1-\half\frac{p^2}{p^2+n_2^2m_0^2}\right)
\frac{1}{2\omega_-(k,1)-\omega_{i_1}(p,n_1)-\omega_{i_2}(p,n_2)+i\e}\right]\nn
.
\ee
with $C=K/(4(2\pi)^4 m \Delta^2)$. Here the reference frame is chosen
where the total transverse momentum vanishes. The energy of both incoming and
scattered photons is $\omega_-(k,1)$, with the photons in the lowest
longitudinal mode
(and where red detuning has been assumed). Since these are low
momentum photons, we have
$\omega_-(k,1)\approx m+k^2/m$.

A simplification may be introduced by considering the effect of mixing
between photons and dipoles, described through the parameter $\eta$.
This mixing affects
the expression in two ways. The first one is through the $\eta$
dependent term that is
explicit in
\pref{Tb}. This term is important when the particle energies are close to the
resonance value $\omega_0$. The other one is the indirect one which
enters through the
particle energy $\omega_{\pm}$. When summing over both the photon and
dipole modes in
the intermediate state this
$\eta$ dependence is less important. It only affects the lowest
longitudinal mode and
only for transverse momenta in a small interval $p\approx \Delta$. We
will neglect this
effect, which can be viewed as higher order in $\eta^2$ and use the
energy expressions
for the uncoupled modes,
\be{energies}
\omega_- &=& \sqrt{p^2+n^2m_0^2} \nn
\omega_+ &=& \omega_0
\ee
(Note, with this redefinition $\omega_-$ is the photon mode also when higher in
energy than the dipole mode.)
We consider now separately contributions from the different modes in
the intermediate
states.

\subsection {Two-dipole intermediate states}
Since $\omega_+$ is independent of $p$ there is no suppression of the
integrand for high
momenta and the integral over $p$ and sum over $n$ are divergent
without the momentum
cutoff. This cutoff we set to $p_{max}$ for the transverse momentum
and $N=p_{max}L/\pi$
for the longitudinal momentum. The connection with the oscillator density is
$\rho=1/l_{osc}^3=(p_{max}/\pi)^3$. Since we assume $L>>l_{osc}$ most
of the contribution
comes from large transverse and longitudinal momenta where the
dimensional reduction no
longer is effective . We note that the only momentum dependence now sits in the
polarization factor, which we approximate by
\be{pol}
1-\half\frac{p^2}{p^2+n^2m_0^2}=1-\half\sin^2\theta\approx \frac 3 4
\ee
where $\theta$ is the angle between the momentum vector and the
$z$-axis. In the last
term we have approximated $\sin^2\theta$ by $1/2$ which is correct
when we (for large
momenta) can treat the longitudinal momentum as a continuous variable.

With these approximations the momentum sum and integral can trivially
be performed, and
the contribution is
\be{Tc}
{\cal T}_{2a} &=& -\frac{9}{32}C^2 \pi N p_{max}^2 \frac{1}{\Delta
\eta^4\omega_0^2} \nn
&=& -\frac{3\pi}{64}\frac{\Lambda}{m^2\Delta}{\cal T}_1
\ee
We note that the sign of this $k$-independent contribution depends on
the sign of
$\Delta$, the detuning parameter.

\subsection {The photon-dipole contribution}
We now consider the terms where one of the intermediate modes is a
photon mode and the
other a dipole mode. The momentum integral is convergent, with the
main contribution
coming from low momenta. For $p<<nm$ we approximate the polarization
factors by 1, and
${\cal T}$ is simplified to
\be{Td}
{\cal T}_{2b} &=& -2\pi C^2\frac{1}{\eta^2\omega_0^2}
\sum_{n}(2+
\frac{1}{4}\delta_{n 1}-
\delta_{n 2}) \nn
&\times & \int_{nm}^{\infty}
dp\left[\frac{p^2}{(p^2-\omega_0^2)^2+\eta^2\omega_0^2}
\;\frac{1}{p+\Delta-m-i\e}\right]
\ee
The leading contribution (in $\Delta/m$) comes from the term $n=1$.
Evaluation of the
momentum integral gives for red detuning $\Delta>0$,
\be{T2bred}
{\cal T}_{2b} &=&
-\frac{9}{128}\frac{1}{(2\pi)^4}\frac{\Lambda^2\eta^5}{m\rho^2\Delta^5}
\nn &=&
-\frac{3}{8}\frac{1}{2\pi}\frac{\Lambda\eta}{\rho\Delta}{\cal T}_1
\ee
and for blue detuning
\be{T2bblue}
{\cal T}_{2b} &=&
\frac{9}{128}\frac{1}{(2\pi)^5}\frac{\Lambda^2\eta^6}{m\rho^2\Delta^6}
-i\frac{9}{512}\frac{1}{(2\pi)^4}\frac{\Lambda^2\eta^6}{m\rho^2
\Delta^6}\nn &=&
\frac{3}{8}\frac{1}{(2\pi)^2}(1-i\frac{\pi}{2})\frac{\Lambda\eta^2}{\rho\Delta^2
}{\cal
T}_1
\ee
  One should note the
difference between red detuning ($\Delta>0$) and blue detuning
($\Delta<0$). In the
latter case the integration path passes a pole which gives rise to an
imaginary part.
That is not the case for red detuning.
\subsection {Two-photon intermediate states}
With two photons in the intermediate state the main contribution
comes from the term
$n_1=n_2=1$ which has a pole at $p^2=k^2$. The low momentum approximation,
$\omega_-(p^2)\approx m+p^2/2m$ can be used, and the expression for
${\cal T}$ simplifies
to
\be{T2c}
{\cal T}_{2c} &=& -\frac{9}{4}\pi C^2 m^3 \int_{0}^{\infty}
dp\,p\frac{1}{((p^2-2m\Delta)^2+\eta^2\omega_0^2)^2}
\;\frac{1}{p^2-k^2-i\e}
\ee
This gives as leading terms for red detuning
  \be{T2cred}
{\cal T}_{2c} &=&
\frac{9}{1024}\frac{1}{(2\pi)^5}\frac{\Lambda^2\eta^8}{m\rho^2\Delta^8}
(\log\frac{k^2}{2m\Delta}-i\pi)+\frac{9}{512}\frac{1}{(2\pi)^4}
\frac{\Lambda^2\eta^5}{m\rho^2\Delta^5}
\nn &=&
\left(\frac{3}{64}\frac{1}{(2\pi)^2}\frac{\Lambda\eta^4}{\rho\Delta^4}
(\log\frac{k^2}{2m\Delta}-i\pi)+\frac{3}{32}\frac{1}{(2\pi)}
\frac{\Lambda\eta}{\rho\Delta}\right){\cal T}_1
\ee
For blue detuning the expression is
\be{T2cblue}
{\cal T}_{2c} &=&
\frac{9}{1024}\frac{1}{(2\pi)^5}\frac{\Lambda^2\eta^8}{m\rho^2\Delta^8}
(\log\frac{k^2}{2m|\Delta|}-i\pi)-\frac{9}{256}\frac{1}{(2\pi)^5}
\frac{\Lambda^2\eta^6}{m\rho^2\Delta^6}
\nn &=&
\left(\frac{3}{64}\frac{1}{(2\pi)^2}\frac{\Lambda\eta^4}{\rho\Delta^4}
(\log\frac{k^2}{2m|\Delta|}-i\pi)-\frac{3}{16}\frac{1}{(2\pi)}
\frac{\Lambda\eta^2}{\rho\Delta^2}\right){\cal T}_1
\ee
The pole at $p^2=k^2$ gives as expected a term that depends logarithmically on
$k^2$.

\subsection{The full expression}

When adding the contributions we find for $\Delta>0$,
  \be{T2totred}
{\cal T}_{2}
&=&\left(\frac{3}{64}\frac{1}{(2\pi)^2}\frac{\Lambda\eta^4}{\rho\Delta^4}
(\log\frac{k^2}{2m\Delta}-i\pi)-\frac{3\pi}{64}\frac{\Lambda}{m^2\Delta}-\frac{9
}{32}\frac{1}{(2\pi)}
\frac{\Lambda\eta}{\rho\Delta}\right){\cal T}_1\nn
&\approx&\left(\frac{3}{64}\frac{1}{(2\pi)^2}\frac{\Lambda\eta^4}{\rho\Delta^4}
(\log\frac{k^2}{2m\Delta}-i\pi)-\frac{3\pi}{64}\frac{\Lambda}{m^2\Delta}\right)
{\cal T}_1
\ee
where in the last expression we have assumed $\eta<<\rho/m^2$ and
have left out the
small term. For
$\Delta<0$ we find
  \be{T2totblue}
{\cal T}_{2}
&=&\left(\frac{3}{64}\frac{1}{(2\pi)^2}\frac{\Lambda\eta^4}{\rho\Delta^4}
(\log\frac{k^2}{2m|\Delta|}-i\pi)-\frac{3\pi}{64}\frac{\Lambda}{m^2\Delta}+
\frac{3}{16}\frac{1}{(2\pi)^2}(1+i\pi)\frac{\Lambda\eta^2}{\rho\Delta^2}
\right){\cal T}_1  \nn
&\approx&\left(\frac{3}{64}\frac{1}{(2\pi)^2}
\frac{\Lambda\eta^4}{\rho\Delta^4}
(\log\frac{k^2}{2m|\Delta|}-i\pi)-\frac{3\pi}{64}\frac{\lambda}{m^2\Delta}+
i\frac{3}{32}\frac{1}{2\pi}\frac{\Lambda\eta^2}{\rho\Delta^2} \right)
{\cal T}_1
\ee
We note that whereas the expression for red detuning agrees with the
expected form for a
renormalized delta function interaction, the expression for blue
detuning contains an
additional imaginary part. This is small compared with the constant
real part, but has
nevertheless some significance. As noted above it arises from the
pole in the energy
factor in the case where one photon and one dipole is present in the
intermediate state.
It is  therefore related to the possibility of real scattering of the two
photons into a photon and a dipole excitation, as discussed in the text.

\vspace{1.5cm}


\begin{thebibliography}{99}
\bibitem{Einstein-A}
The detuning $\Delta$ must be much larger than the natural
linewidth, in order to avoid any $real$ transitions. In particular,
it is necessary that $\Delta \gg A$, where $A$ is the Einstein $A$
coefficient for the rate
of spontaneous emission. Since $A \sim {\omega_{0}}^{3}$, this
restriction is not
important for the case of microwave transitions of the Rydberg atoms, but could
become important for the case of optical transitions of alkali atoms.
Boyd \cite{Boyd} in Eq.~(3.5.39) gives
\begin{eqnarray}
A=\frac{4{\omega_{0}}^{3}\mu_e^2}{\hbar c^3}
=\frac{4\pi c\alpha_{e}
\lambda_{C}f_{osc}}{{\lambda_{0}}^{2}}
= 0.67 ~
\frac{f_{osc}}{{\lambda_{0}}^{2}}
~{\rm (cgs~units).}
\end{eqnarray}
For the case of the $n=51 \rightarrow 50$ Rydberg
atom transition, one finds that $A\simeq 93 {\rm~s}^{-1}$, so that
the detuning must be much
greater than 15 Hz, which is not much of a restriction. For the case
of rubidium
atoms in their ground atoms excited near the resonance wavelength of
$\lambda_{0} \approx 780 {\rm~nm}$, one finds that $A \simeq 1.1 \times
10^{8} {\rm~s}^{-1}$, so that a detuning of much greater than 18 MHz is
required. We shall later use the value of $\Delta = 2\pi \times 280 {\rm~MHz}$,
which satisfies this requirement.
\end{thebibliography}
\end{document}